\documentclass[prb,twocolumn,floatfix]{revtex4}

\usepackage{graphicx}
\usepackage{amsmath}
\usepackage{amssymb}
\usepackage[colorlinks=true,citecolor=blue,linkcolor=blue]{hyperref}
\usepackage{amsfonts}
\usepackage{color,xcolor}
\usepackage{epstopdf}
\usepackage{braket}
\usepackage{bm}
\usepackage{bbold}

\renewcommand{\vec}[1]{\ensuremath{\boldsymbol{#1}}}

\begin{document}

\title{Excitons and trions in monolayer transition metal dichalcogenides: a comparitive study between the multi-band model and the quadratic single-band model}
\date{\today}
\author{M. Van der Donck}
\email{matthias.vanderdonck@uantwerpen.be}
\affiliation{Department of Physics, University of Antwerp, Groenenborgerlaan 171, B-2020 Antwerp, Belgium}
\author{M. Zarenia}
\email{mohammad.zarenia@uantwerpen.be}
\affiliation{Department of Physics, University of Antwerp, Groenenborgerlaan 171, B-2020 Antwerp, Belgium}
\author{F. M. Peeters}
\email{francois.peeters@uantwerpen.be}
\affiliation{Department of Physics, University of Antwerp, Groenenborgerlaan 171, B-2020 Antwerp, Belgium}

\begin{abstract}
The electronic and structural properties of excitons and trions in monolayer transition metal dichalcogenides are investigated using both a multi-band and a single-band model. In the multi-band model we construct the excitonic Hamiltonian in the product base of the single-particle states at the conduction and valence band edges. We decouple the corresponding energy eigenvalue equation and solve the resulting differential equation self-consistently, using the finite element method (FEM), to determine the energy eigenvalues and the wave functions. As a comparison, we also consider the simple single-band model which is often used in numerical studies. We solve the energy eigenvalue equation using the FEM as well as with the stochastic variational method (SVM) in which a variational wave function is expanded in a basis of a large number of correlated Gaussians. We find good agreement between the results of both methods as well as with other theoretical works for excitons and we also compare with available experimental data. For trions the agreement between both methods is not as good due to our neglect of angular correlations when using the FEM. Finally, when comparing the two models, we see that the presence of the valence bands in the muti-band model leads to differences with the single-band model when (interband) interactions are strong.

\end{abstract}

\maketitle

\section{Introduction}
Two dimensional (2D) atomically thin materials, such as graphene \cite{novo1} and transition-metal dichalcogenide (TMD) monolayers as MoS$_2$, MoSe$_2$, WS$_2$, WSe$_2$, WTe$_2$, etc. \cite{mak1,mak2,expT3,zeng,splendiani,cao}, have attracted the attention of the condensed matter community because they display new fundamental physics and because their remarkable electronic properties are expected to be important for future applications in electronics and optics. In contrast to graphene, which has a gapless and linear spectrum \cite{review}, inversion symmetry breaking in TMD monolayers leads to the formation of a direct band gap which is located at the two inequivalent $K$ and $K'$ valleys at the corners of the first Brillouin zone. Moreover, monolayer TMDs have an intrinsic spin-orbit coupling resulting in a splitting of the valence bands with opposite spins \cite{theory1}. These properties have propelled initial efforts to demonstrate valley polarization and related novelties for device applications based on valleytronics \cite{mak2,zeng,cao}.

It was realized that for ultrathin semiconductors, the dielectric environment plays a crucial role and influences the effective strength of the Coulomb potentials inside a semiconductor layer \cite{mismatch}. Such long-range interactions become stronger as the thickness of the semiconductor layer decreases which allows the formation of tightly bound \emph{excitons} (electron-hole pairs). This enhanced Coulomb interaction leads to exciton binding energies of the order 0.5-1 eV  in TMD monolayers which are one to two orders of magnitude larger than excitons in typical semiconductors, which have been investigated for more than half a century \cite{elliot,kulak, francoisE,francoisT,expT0}. Recent photoluminescence experiments in monolayer MoS$_2$, MoSe$_2$, WS$_2$, and WSe$_2$ confirmed the existence of excitonic states that are localized in the band gap \cite{mak3,chernikov,he,sallen,korn}. Furthermore, a few theoretical works pertinent to the excitonic absorption spectrum of these materials have appeared recently \cite{exctheory1,exctheory2,exctheory3}.

In addition to excitons, \emph{trions} have also been observed in TMDs. Trions are charged excitons that consist of an electron (e) or a hole (h) bound to an exciton (X). Since the first prediction of the existence of two kinds of trions (X$^+$ and X$^-$) in bulk semiconductors in 1958 \cite{lampert}, there have been many theoretical \cite{francoisE,francoisT,munschy} and experimental \cite{expT3,expT0,expT1,expT2,expT4} studies on trions in different systems such as e.g. semiconductor quantum wells (for example see Refs. [\onlinecite{francoisE,francoisT,expT0}]). Recent spectroscopic measurements on monolayer MoS$_2$ and WSe$_2$ have demonstrated the existence of tightly bound trions \cite{mak3,lui,trionNew} with unprecedented binding energies, i.e. 20-30 meV, which compares with 0.5-3 meV for trions in GaAs quantum wells \cite{qwell}.  The binding energy of trions in TMDs was recently calculated by Berkelbach {\it et al.} \cite{berkelbach}. 

Here we present a theoretical analysis of the electronic and structural properties of excitons and trions in monolayer TMDs using two different models. In the multi-band model we construct the excitonic Hamiltonian, including the effect of spin-orbit coupling, in the product base of the single-particle states at the conduction and valence band edges \cite{decouple1}. Such a model was used earlier to describe excitonic superfluidity in double-layer graphene \cite{decouple2} and in 2D TMDs \cite{decouple4}. We decouple the corresponding energy eigenvalue equation and solve the resulting differential equation self-consistently, using the finite element method (FEM). An import advantage of this approach is that it allows us to readily obtain the excited excitonic states.

As a comparison, we also consider the simple single-band model which is often used in numerical studies and we solve the eigenvalue equation using the FEM as well as the stochastic variational method (SVM) using a correlated Gaussian basis \cite{svm1,svm2}. The SVM was successfully used to describe the binding energy of excitons, trions, and biexcitons in semiconductor quantum wells \cite{francoisE} and even their magnetic field dependence \cite{francoisT}. Recently the SVM was used to calculate the binding energies of excitons, trions, and biexcitons in TMD monolayers \cite{analytic}. It was demonstrated that the theoretical results of this work are in good agreement with experiments and other theoretical results. In our work, we employ this approach to calculate the binding energy and wave function of excitons and trions in different monolayer TMDs and compare the results with the FEM results for the single-band model and the multi-band model.

We also present a detailed comparison with available theoretical and experimental results for the binding energy of excitons and trions in TMDs and we demonstrate that the multi-band model can properly describe the experimental absorbance spectrum.

Our paper is organized as follows. In Sec. \ref{sec:Model} we present the outline of the multi-band model and the single-band model together with an explanation of the SVM. The numerical results for both excitons and trions, in both the multi-band and the single-band model, are discussed in Sec. \ref{sec:Results} and a comparison is made between the FEM and the SVM results. In Sec. \ref{sec:Summary and conclusion} we summarize the main conclusions.

\section{Model}
\label{sec:Model}

\subsection{Multi-band model}

Excitons and trions are many-body systems, requiring the use of quantum field theory. However, these excitonic systems can be well approximated by treating them as few-body systems. We start from the effective low-energy single-electron Hamiltonian\cite{theory1} in the basis $\mathcal{B}^e=\{\ket{\phi^e_c},\ket{\phi^e_v}\}$ spanning the 2D Hilbert space $\mathcal{H}^e$, with $\ket{\phi^e_c}$ and $\ket{\phi^e_v}$ the atomic orbital states at the conduction $(c)$ and valence $(v)$ band edge, respectively:
\begin{equation}
\label{singelhame}
H^e_{s,\tau}(\vec{k}) = at(\tau k_x\sigma_x+k_y\sigma_y)+\frac{\Delta}{2}\sigma_z+\lambda s\tau\frac{I_2-\sigma_z}{2},
\end{equation}
where $\sigma_i$ ($i=x,y,z$) are Pauli matrices, $I_2$ is the two by two identity matrix, $a$ the lattice constant, $t$ the hopping parameter, $\tau=\pm1$ the valley index, $s=\pm1$ the spin index, $\Delta$ the band gap, and $\lambda$ the spin-orbit coupling strength leading to a spin splitting of $2\lambda$ at the valence band edge. Since a hole with wave vector $\vec{k}$, spin $s$, and valley index $\tau$ can be described as the absence of an electron with opposite wave vector, spin, and valley index, the single-hole Hamiltonian can immediately be obtained from the single-electron Hamiltonian as $\hat{H}^h_{s,\tau}(\vec{k})=-\hat{H}^e_{-s,-\tau}(-\vec{k})$, and the eigenstates of this Hamiltonian span the Hilbert space $\mathcal{H}^h$. The total Hamiltonian of $N$ electrons and $M$ holes is given by the sum of the separate single-particle Hamiltonians and the corresponding Hilbert space is given by the product space of all the Hilbert spaces spanned by the eigenstates of the separate Hamiltonians, $\mathcal{H}^{tot}=\mathcal{H}^e_1\otimes\ldots\otimes\mathcal{H}^e_N\otimes\mathcal{H}^h_1\otimes\ldots\otimes\mathcal{H}^h_M$, and has dimension $2^{N+M}$. The most straightforward set of basis states spanning this total Hilbert space is given by the set of all the possible combinations of tensor products of the atomic orbital states of the individual particles at the conduction and valence band edges. In schematic notation this can be written as $\mathcal{B} = \mathcal{B}^e_1\otimes\ldots\otimes\mathcal{B}^e_N\otimes\mathcal{B}^h_1\otimes\ldots\otimes\mathcal{B}^h_M$, which is a set of $2^{N+M}$ states as is required.

\begin{figure}
\centering
\includegraphics[width=8.5cm]{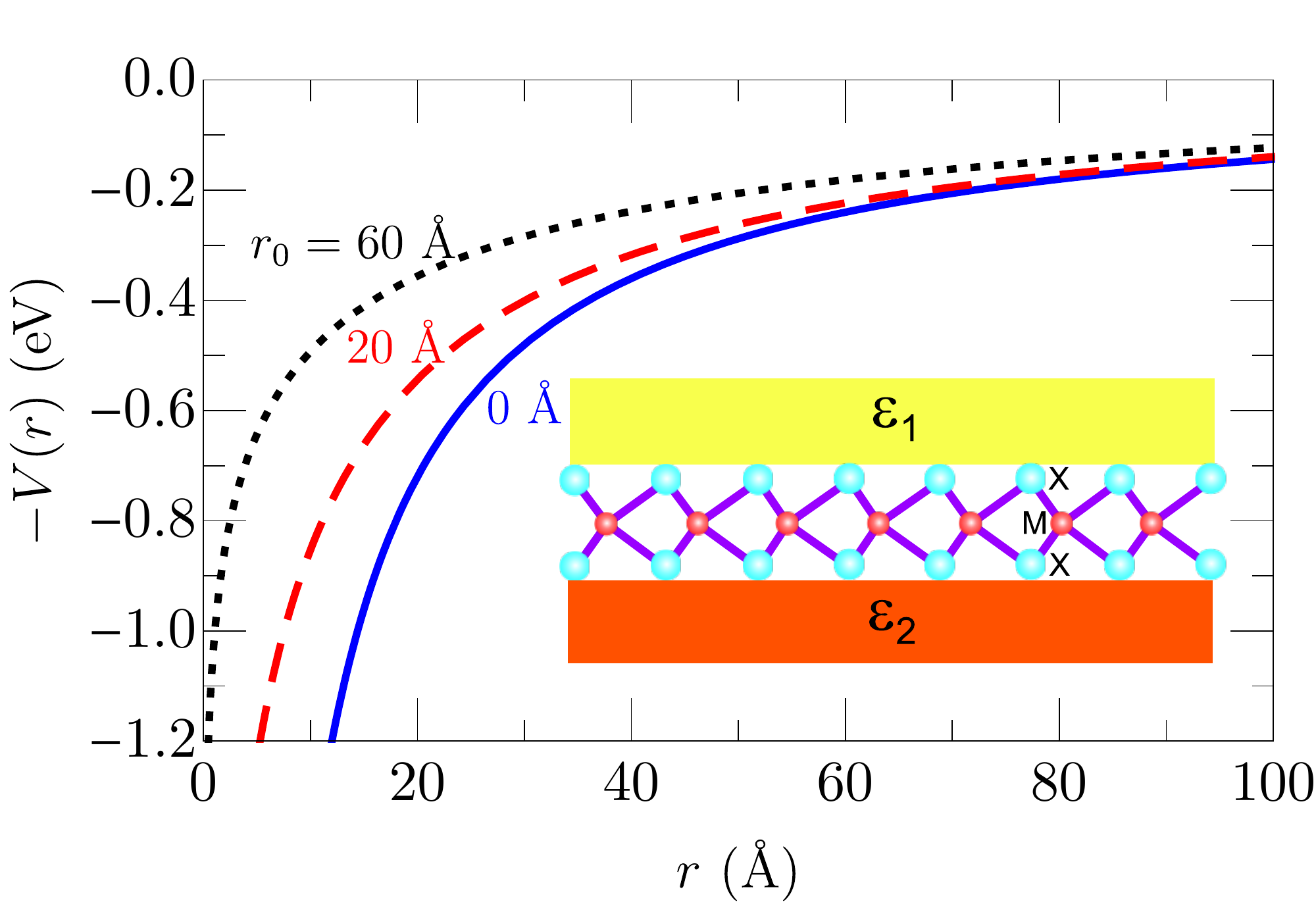}
\caption{(Color online) Interaction potential between a hole and an electron in a monolayer TMD for $\kappa=1$ and with screening length $r_0=0$ \AA\ (solid, blue), $r_0=20$ \AA\ (dashed, red), and $r_0=60$ \AA\ (dotted, black). The inset shows a schematic illustration of a monolayer TMD placed between two dielectrica.}
\label{fig:screening}
\end{figure}

Interactions between the different particles can now be added and will enter in the total excitonic Hamiltonian as
\begin{equation}
\label{inter}
\sum_{i<j}^{N+M}\text{sgn}(q_iq_j)V(|\vec{r}_i-\vec{r}_j|)I_{2^{N+M}},
\end{equation}
The TMD monolayer is surrounded by a dielectric with a dielectric constant different from that of the TMD, as shown in the inset of Fig. \ref{fig:screening}. It is well-known that this leads to a particular screening of the $1/r$ Coulomb potential such that the interaction potential $V_{ij}$ is now given by the 2D screened potential\cite{screening1,screening2,screening3}
\begin{equation}
\label{interpot}
V(r_{ij}) = \frac{e^2}{4\pi\kappa\varepsilon_0}\frac{\pi}{2r_0}\left[H_0\left(\frac{r_{ij}}{r_0}\right)-Y_0\left(\frac{r_{ij}}{r_0}\right)\right],
\end{equation}
with $r_{ij}=|\vec{r}_i-\vec{r}_j|$, where $Y_0$ and $H_0$ are the Bessel function of the second kind and the Struve function, respectively, with $\kappa=(\varepsilon_1+\varepsilon_2)/2$ where $\varepsilon_{1(2)}$ is the dielectric constant of the environment above (below) the TMD monolayer, and with $r_0=2\pi\chi_{2\text{D}}/\kappa$ the screening length where $\chi_{2\text{D}}$ is the 2D polarizability of the TMD. In this work we will always consider TMDs placed on a substrate with a dielectric constant $\varepsilon_2=\varepsilon_r$ and with vacuum on top, i.e. $\varepsilon_1=1$. The interaction potential is shown in Fig. \ref{fig:screening} for different screening lengths. For $r_0=0$ this potential reduces to the bare Coulomb potential $V(r_{ij})=e^2/(4\pi\kappa\varepsilon_0r_{ij})$. Increasing the screening length leads to a decrease in the short-range interaction strength while the long-range interaction strength is unaffected. For very large screening lengths $r_0\rightarrow\infty$ the interaction potential becomes logarithmic, i.e. $V(r_{ij})=e^2/(4\pi\kappa\varepsilon_0r_0)\text{ln}(r_0/r_{ij})$.

\subsubsection{Exciton}

The exciton Hamiltonian is constructed in the basis $\mathcal{B}^{exc}=\{\ket{\phi^e_c}\otimes\ket{\phi^h_c},\ket{\phi^e_c}\otimes\ket{\phi^h_v},\ket{\phi^e_v}\otimes\ket{\phi^h_c},\ket{\phi^e_v}\otimes\ket{\phi^h_v}\}$. As an example, the non-interacting matrix element between the first and second basis state is given by
\begin{equation}
\label{H12}
\begin{split}
H_{1,2}^{exc} &= \left(\bra{\phi^e_c}\otimes\bra{\phi^h_c}\right)\Big(\hat{H}^e_{s^e,\tau^e}(\vec{k}^e)\otimes\mathbb{1} \\
&\quad\quad\quad+\mathbb{1}\otimes\hat{H}^h_{s^h,\tau^h}(\vec{k}^h)\Big)\left(\ket{\phi^e_c}\otimes\ket{\phi^h_v}\right) \\
&= at(-\tau^hk_x^h-ik_y^h),
\end{split}
\end{equation}
where the orthonormality of the basis functions was used. The other matrix elements can be calculated in a similar way and this gives the total exciton Hamiltonian
\begin{widetext}
\begin{equation}
\label{hamtot}
H^{exc}_{\alpha}(\vec{k}^e,\vec{k}^h,r_{eh}) =
\begin{pmatrix}
-V(r_{eh}) & at(-\tau^hk_x^h-ik_y^h) & at(\tau^ek_x^e-ik_y^e) & 0 \\
at(-\tau^hk_x^h+ik_y^h) & \Delta-\lambda s^h\tau^h-V(r_{eh}) & 0 & at(\tau^ek_x^e-ik_y^e) \\
at(\tau^ek_x^e+ik_y^e) & 0 & -\Delta+\lambda s^e\tau^e-V(r_{eh}) & at(-\tau^hk_x^h-ik_y^h) \\
0 & at(\tau^ek_x^e+ik_y^e) & at(-\tau^hk_x^h+ik_y^h) & \lambda(s^e\tau^e-s^h\tau^h)-V(r_{eh})
\end{pmatrix},
\end{equation}
\end{widetext}
where the interaction terms have now been added and where $\alpha$ is a shorthand notation for $s^e,\tau^e,s^h,\tau^h$. The eigenvalue problem for this Hamiltonian,
\begin{equation}
\label{eigen}
H^{exc}_{\alpha}(\vec{k}^e,\vec{k}^h,r_{eh})\ket{\Psi^{exc}_{\alpha}} = E^{exc}_{\alpha}(\vec{k}^e,\vec{k}^h)\ket{\Psi^{exc}_{\alpha}},
\end{equation}
defines the exciton energy $E^{exc}_{\alpha}(\vec{k}^e,\vec{k}^h)$ and the exciton eigenstate $\ket{\Psi^{exc}_{\alpha}}=\left(\ket{\phi^{e,h}_{c,c}},\ket{\phi^{e,h}_{c,v}},\ket{\phi^{e,h}_{v,c}},\ket{\phi^{e,h}_{v,v}}\right)^T$, where the subscript $\alpha$ and the superscript $exc$ have been dropped in the right hand side for notational clarity. The above eigenvalue problem is a matrix equation which can, following a procedure analogous to earlier works \cite{decouple1,decouple2,decouple3,decouple4}, be decoupled to a single equation, as shown in Appendix \ref{sec:appA}. For $s$-state excitons created by exciting charge carriers with \textit{circularly polarized light}, implying that the electrons and holes are created in a single valley and therefore have both opposite spin index as well as opposite valley index, this equation reduces to
\begin{equation}
\label{difvgl}
\begin{split}
\bigg(&-\frac{2a^2t^2}{E^{exc}_{\alpha}+V(r)}\nabla^2_{\vec{r}}-2a^2t^2\left(\frac{\partial}{\partial r}\frac{1}{E^{exc}_{\alpha}+V(r)}\right)\frac{\partial}{\partial r} \\
&-V(r)+\Delta_{s^h,\tau^h}\bigg)\phi_{c,v}^{e,h}(r) = E^{exc}_{\alpha}\phi_{c,v}^{e,h}(r),
\end{split}
\end{equation}
with $\Delta_{s^h,\tau^h}=\Delta-\lambda s^h\tau^h$ the effective band gap. When exciting the charge carriers with \textit{linearly polarized light}, implying that electrons and holes are created in both valleys, there will also be electrons and holes with the same valley index, however these will also have the same spin index and as such the equation above still holds. Therefore, in general, two particles can be excited simultaneously if they have the same value of $s\tau$. This equation is a differential eigenvalue equation, which we solve with the FEM, with the additional complication of the eigenvalue appearing in the left hand side as well. Therefore we have to solve this equation self-consistently by choosing an initial value for $E^{exc}_{\alpha}$ and inserting it in the left hand side and numerically calculating the corresponding eigenvalue in the right hand side. This newly calculated eigenvalue is subsequently used in the left hand side to calculate a new eigenvalue. This is repeated until convergence is reached. When the exciton energy $E^{exc}_{\alpha}$ is calculated, the binding energy is also known as it is given by $E_{b,\alpha}^{exc}=\Delta_{s^h,\tau^h}-E^{exc}_{\alpha}$.

\subsubsection{Trion}

In this paper we will only consider negative trions, as opposed to positive trions. However, the two kinds of trions exhibit very similar properties and the procedure presented below can very easily be modified to describe positive trions. We construct the trion Hamiltonian in the basis $\mathcal{B}^{tri}=\{\mathcal{B}^{exc}\otimes\ket{\phi^{e_2}_c},\mathcal{B}^{exc}\otimes\ket{\phi^{e_2}_v}\}$, in which it can be written as
\begin{equation}
\label{hamtri}
H^{tri}_{\alpha,s^{e_2},\tau^{e_2}} =
\begin{pmatrix}
H^{exc}_{\alpha}+V_1 & \mathcal{O}_{e_2} \\
\mathcal{O}^{\dag}_{e_2} & H^{exc}_{\alpha}+V_2^{s^{e_2},\tau^{e_2}}
\end{pmatrix},
\end{equation}
with $H^{exc}_{\alpha}$ the exciton Hamiltonian \eqref{hamtot} and with
\begin{align}
&\mathcal{O}_{e_2} = at\left(\tau^{e_2}k_x^{e_2}-ik_y^{e_2}\right)I_4, \label{O3} \\
&V_1 = \left(\frac{\Delta}{2}-V(r_{he_2})+V(r_{e_1e_2})\right)I_4, \label{V1} \\
&V_2^{s^{e_2},\tau^{e_2}} = \left(-\frac{\Delta}{2}+\lambda s^{e_2}\tau^{e_2}-V(r_{he_2})+V(r_{e_1e_2})\right)I_4. \label{V2}
\end{align}
From the second equation of the eigenvalue problem $H^{tri}_{\alpha,s^{e_2},\tau^{e_2}}\ket{\Psi^{tri}_{\alpha,s^{e_2},\tau^{e_2}}}=E^{tri}_{\alpha,s^{e_2},\tau^{e_2}}\ket{\Psi^{tri}_{\alpha,s^{e_2},\tau^{e_2}}}$, it follows that
\begin{equation}
\label{reltri}
\ket{\Psi^{e_2}_v} \approx \left(E^{tri}_{\alpha,s^{e_2},\tau^{e_2}}I_4-V_2^{s^{e_2},\tau^{e_2}}-D^{exc}_{\alpha}\right)^{-1}\mathcal{O}^{\dag}_{e_2}\ket{\Psi^{e_2}_c},
\end{equation}
with $\ket{\Psi^{e_2}_c}=\left(\ket{\phi^{e_1,h,e_2}_{c,c,c}},\ket{\phi^{e_1,h,e_2}_{c,v,c}},\ket{\phi^{e_1,h,e_2}_{v,c,c}},\ket{\phi^{e_1,h,e_2}_{v,v,c}}\right)^T$, $\ket{\Psi^{e_2}_v}=\left(\ket{\phi^{e_1,h,e_2}_{c,c,v}},\ket{\phi^{e_1,h,e_2}_{c,v,v}},\ket{\phi^{e_1,h,e_2}_{v,c,v}},\ket{\phi^{e_1,h,e_2}_{v,v,v}}\right)^T$, and with $D^{exc}_{\alpha}$ the four by four diagonal matrix containing the diagonal elements of the exciton Hamiltonian \eqref{hamtot}. In this approximation, the kinetic energy of the first electron and the hole is assumed to be small compared to the band gap and the trion energy. Inserting the above relation in the first equation of the eigenvalue problem we find
\begin{equation}
\label{eigtri}
\begin{split}
&\Big(H^{exc}_{\alpha}+\mathcal{O}_{e_2}\left(E^{tri}_{\alpha,s^{e_2},\tau^{e_2}}I_4-V_2^{s^{e_2},\tau^{e_2}}-D^{exc}_{\alpha}\right)^{-1}\mathcal{O}^{\dag}_{e_2} \\
&\hspace{6pt}+V_1\Big)\ket{\Psi^{e_2}_c} = E^{tri}_{\alpha,s^{e_2},\tau^{e_2}}\ket{\Psi^{e_2}_c}.
\end{split}
\end{equation}
This eigenvalue problem is similar to the exciton eigenvalue problem \eqref{eigen}, but now with additional terms on the diagonal. Therefore, this four by four matrix equation can be decoupled in a similar fashion as described in Appendix \ref{sec:appA} for excitons and the resulting differential equation is shown in Eq. \eqref{tridif}. In the derivation of this differential equation, we assumed that the wave function is independent of the angular coordinates $\varphi_{e_1h}$ and $\varphi_{he_2}$. Therefore, as an approximation we take $\varphi_{e_1h}=\varphi_{he_2}$ such that we have $|\vec{r}_{e_1h}+\vec{r}_{he_2}|= r_{e_1h}+r_{he_2}$. Equation \eqref{tridif} again has to be solved self-consistently to determine the trion energy $E^{tri}_{\beta}$ and the component $\phi_{c,v,c}^{e_1,h,e_2}(r_{e_1h},r_{he_2})$ of the trion wave function. The other components of the wave function can be determined from Eq. \eqref{reltri} and equations which are analogous to Eqs. \eqref{rel1} and \eqref{rel2} and which can be found by decoupling the eigenvalue problem. The trion binding energy is given by $E_{b,\beta}^{tri}=\Delta/2+E^{exc}_{\alpha}-E^{tri}_{\beta}$.

\subsection{Single-band model}

The Hamiltonian for an $N$-particle excitonic system can also be approximately written in the form
\begin{equation}
\label{ham}
H = \sum_{i=1}^N\frac{\hbar^2k_i^2}{2m_i}+\sum_{i<j}^N\text{sgn}(q_iq_j)V(|\vec{r}_i-\vec{r}_j|),
\end{equation}
with $q_i$ and $m_i$ the charge and effective mass of particle $i$ and where $V(|\vec{r}_i-\vec{r}_j|)$ is again given by Eq. \eqref{interpot}. Here, we assumed that the electron and hole bands are isotropic and parabolic, which is a good approximation for the low-energy spectrum of the considered TMDs. However, the above form of the Hamiltonian implies that both the electron and hole single-particle states form a single parabolic band. The corresponding energy eigenvalue equation is given by
\begin{equation}
\label{schrodexc}
\left(-\frac{\hbar^2}{2\mu}\nabla_{\vec{r}}^2-V(r)\right)\psi^{exc}(r) = E^{exc}\psi^{exc}(r),
\end{equation}
and
\begin{equation}
\label{trischrod}
\begin{split}
\bigg(&-\frac{\hbar^2}{2\mu}\left(\nabla_{\vec{r}_{e_1h}}^2+\nabla_{\vec{r}_{he_2}}^2\right)+\frac{\hbar^2}{m_h}\vec{\nabla_{\vec{r}_{e_1h}}}.\vec{\nabla_{\vec{r}_{he_2}}} \\
&-V(r_{12})-V(r_{23})+V(r_{12}+r_{23})\bigg)\psi^{tri}(r_{12},r_{23}) \\
&= E^{tri}\psi^{tri}(r_{12},r_{23}),
\end{split}
\end{equation}
for $s$-state excitons and negative trions, respectively, and with $\mu=(1/m_e+1/m_h)^{-1}$ the reduced mass. Here, we again transformed to center of mass and relative coordinates and take $\vec{K}=\vec{0}$. These differential equations can be solved directly with the FEM, although for the trion differential equation we have to make the approximation that the wave function is independent of the relative angular coordinates and take $|\vec{r}_{e_1h}+\vec{r}_{he_2}|= r_{e_1h}+r_{he_2}$.

However, it is possible to determine the ground state energy and wave function of the above Hamiltonian including all the angular correlations. In order to do this, we follow Ref. [\onlinecite{analytic}] and employ the SVM, which allows to solve this Hamiltonian quasi-exactly. We expand the many-particle wave function $\Psi(\vec{r}_1,\ldots,\vec{r}_N)$ in a basis of given size $K$:
\begin{equation}
\label{basisexp}
\Psi_{M_L,S,M_S}(\vec{r}_1,\ldots,\vec{r}_N) = \sum_{n=1}^Kc_n\varphi_{M_L,S,M_S}^{n}(\vec{r}_1,\ldots,\vec{r}_N),
\end{equation}
where the basis functions are taken as correlated Gaussians:
\begin{equation}
\label{corrgaus}
\begin{split}
&\varphi_{M_L,S,M_S}^{n}(\vec{r}_1,\ldots,\vec{r}_N) = \\
&\mathcal{A}\left(\Bigg(\prod_{j=1}^N\xi_{m_j^n}(\vec{r}_j)\Bigg)e^{-\frac{1}{2}\sum_{i,j=1}^NA_{ij}^n\vec{r}_i.\vec{r}_j}\chi^n_{S,M_S}\right),
\end{split}
\end{equation}
with $\xi_m(\vec{r})=(x+\text{sgn}(m)iy)^{|m|}$. The matrix elements $A_{ij}^n$ are the variational parameters and form a symmetric and positive definite matrix $A^n$. $\chi^n_{S,M_S}$ is the total spin state of the excitonic system corresponding to the total spin $S$ and $z$-component of the spin $M_S$, which are conserved quantities. This total spin state is obtained by adding step by step single-particle spin states. Therefore, multiple total spin states belonging to the same $S$ and $M_S$ value are possible, as these can be obtained by different intermediate spin states. The integers $m_j^n$ satisfy the relation $\sum_{j=1}^Nm_j^n=M_L$ with $M_L$ the $z$-component of the total angular momentum, which is also conserved. For the exciton we consider the $(S,M_S)=(0,0)$ singlet state and for the trion we consider the $(S,M_S)=(1/2,1/2)$ doublet state. Furthermore, we always take $M_L=0$, which is the lowest energy state. Finally, $\mathcal{A}$ is the antisymmetrization operator for the indistinguishable particles. The matrix elements of the different terms of the Hamiltonian between these basis functions can be calculated analytically\cite{analytic}.

To find the best energy value, we randomly generate a matrix $A^n$, integers $m_j^n$ and a spin function $\chi^n_{S,M_S}$ multiple times. The wave function with the set of parameters that gives the lowest energy is then retained as a basis function, and we now have a basis of dimension $K=1$. Subsequently, we again randomly generate a set of parameters and calculate the energy value in the $K=2$ basis consisting of our previously determined basis function and the new trial basis function. This is repeated multiple times and the trial function that gives the lowest energy value is then retained as the second basis function. Following this procedure, each addition of a new basis function assures a lower variational energy value and we keep increasing our basis size until we reach convergence of the energy value. This procedure is explained in more detail in Ref. [\onlinecite{svm1}].

In this model, the binding energies for excitons and trions are, respectively, given by $E_b^{exc} = -E^{exc}$ and $E_b^{tri} = E^{exc}-E^{tri}$, where $E^{exc}$ and $E^{tri}$ are the exciton and trion energy, respectively.

\section{Results}
\label{sec:Results}

\begin{table}
\centering
\caption{Lattice constants \cite{theory1}, hopping parameters \cite{theory1}, band gaps \cite{theory1}, spin splittings \cite{spinsplit}, charge carrier masses \cite{berkelbach}, and screening lengths \cite{berkelbach} for different TMD materials suspended in vacuum.}
\begin{tabular}{c c c c c c c}
\hline
\hline
 & $a$ (\AA) & $t$ (eV) & $\Delta$ (eV) & $2\lambda$ (eV)& $m$ ($m_0$) & $r_0$ (\AA) \\
\hline
\hline
Mo$\text{S}_2$ & 3.193 & 1.10 & 1.66 & 0.15 & 0.50 & 41.47 \\
\hline
MoS$\text{e}_2$ & 3.313 & 0.94 & 1.47 & 0.18 & 0.54 & 51.71 \\
\hline
W$\text{S}_2$ & 3.197 & 1.37 & 1.79 & 0.43 & 0.32 & 37.89 \\
\hline
WS$\text{e}_2$ & 3.310 & 1.19 & 1.60 & 0.46 & 0.34 & 45.11 \\
\hline
\hline
\end{tabular}
\label{table:mattable}
\end{table}

\begin{figure}
\centering
\includegraphics[width=8.5cm]{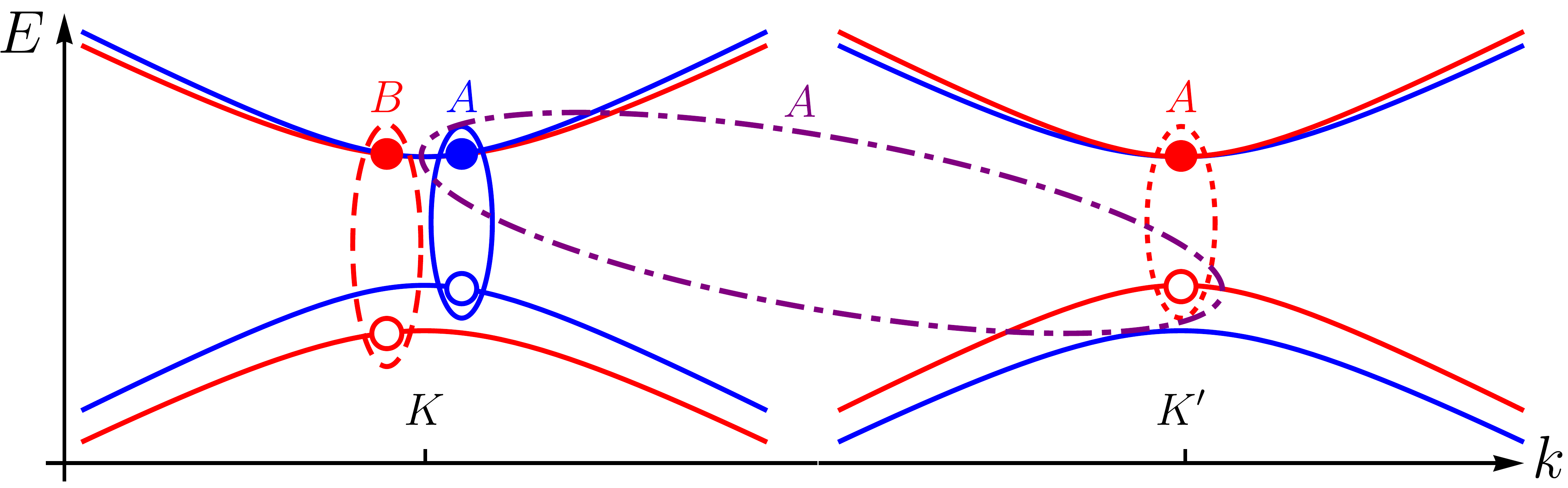}
\caption{(Color online) Schematic representation of the low-energy band structure of 2D TMDs and different kinds of excitons. Blue and red bands are spin up and spin down bands, respectively. The open and closed circles indicate holes and electrons, respectively. The blue solid ellipse and the red dotted ellipse indicate intravalley $A$ excitons in the $K$ and $K'$ valley, respectively. The red dashed ellipse indicates an intravalley $B$ exciton in the $K$ valley. The large purple dot-dashed ellipse indicates an intervalley $A$ exciton.}
\label{fig:schema}
\end{figure}

The material constants for four different TMDs are listed in Table \ref{table:mattable}. Unless specified otherwise, all calculations are done for MoS$_2$ suspended in vacuum. However, when comparing the two models, not all these parameters are independent. The single-electron energy spectrum following from the multi-band Hamiltonian \eqref{singelhame} is given by
\begin{equation}
\label{singelspece}
E_{s,\tau}(\vec{k}) = \frac{\lambda s\tau}{2}\pm\sqrt{a^2t^2k^2+\frac{\Delta_{s,\tau}^2}{4}},
\end{equation}
and is shown schematically in Fig. \ref{fig:schema}. For small $k$ this energy spectrum can be approximated by
\begin{equation}
\label{singelspecapprox}
E_{s,\tau}(\vec{k}) \approx \frac{\lambda\tau s\pm\Delta_{s,\tau}}{2}\pm\frac{\hbar^2k^2}{2m_{s,\tau}},
\end{equation}
with the charge carrier mass given by
\begin{equation}
\label{mass}
m_{s,\tau} = \frac{\hbar^2\Delta_{s,\tau}}{2a^2t^2}.
\end{equation}
Note that this implies that particles with different spin index or with different valley index have a different effective mass. This quadratic approximation is inherent to the single-band model and is therefore often made in numerical studies of excitonic systems such as in Monte Carlo calculations and in SVM calculations. Even at high charge carrier densities of $10^{13}$ cm$^{-2}$ the quadratic dispersion overestimates the full hyperbolic dispersion by an amount of the order of only 0.1-1 meV. However, in the single-band model the energy spectrum in Eq. \eqref{singelspecapprox} is further approximated to a single band and as such a parameter is lost. We have two (effective) parameters ($at$ and $\Delta_{s,\tau}$) in the multi-band model and only one ($m_{s,\tau}$) in the single-band model. We will elaborate more on this in the next subsection.

In both the multi-band model as well as the single-band model, we will calculate the correlation function between two particles $i$ and $j$, defined as
\begin{equation}
\label{corr}
C_{ij}(\vec{r}) = \braket{\Psi|\delta(\vec{r}_i-\vec{r}_j-\vec{r})|\Psi},
\end{equation}
from which we can calculate the probability of finding particles $i$ and $j$ at a distance $r$. For an axisymmetric system, this reduces to
\begin{equation}
\label{prob}
P_{ij}(r) = 2\pi rC_{ij}(r),
\end{equation}
which satisfies
\begin{equation}
\label{norm}
\int_0^{\infty}P_{ij}(r)dr = 1.
\end{equation}
The average distance between particles $i$ and $j$ is then obtained by
\begin{equation}
\label{dist}
\braket{r_{ij}} = \int_0^{\infty}rP_{ij}(r)dr = 2\pi\int_0^{\infty}r^2C_{ij}(r)dr.
\end{equation}

When both the excitonic energy spectrum as well as the wave functions are known we can also calculate the absorbance spectrum using the formula\cite{absorbform}
\begin{equation}
\label{absorbance}
\alpha(\omega) \propto \frac{1}{\omega}\text{Im}\left(\sum_j\frac{|\mathcal{P}_0|^2|\phi_{c,v}^{e,h,j}(0)|^2}{E_j-\hbar\omega-i\gamma}\right),
\end{equation}
with $E_j$ the exciton energy, $\phi_{c,v}^{e,h,j}$ the corresponding dominant component of the exciton wave function, $\hbar\omega$ the photon energy, $\gamma$ the broadening of the peaks and where $\mathcal{P}_0=2m_{s,\tau}at/\hbar$ is the coupling strength with optical fields of circular polarization evaluated at the band edges\cite{theory1}.

\subsection{Exciton}

\begin{figure}
\centering
\includegraphics[width=8.5cm]{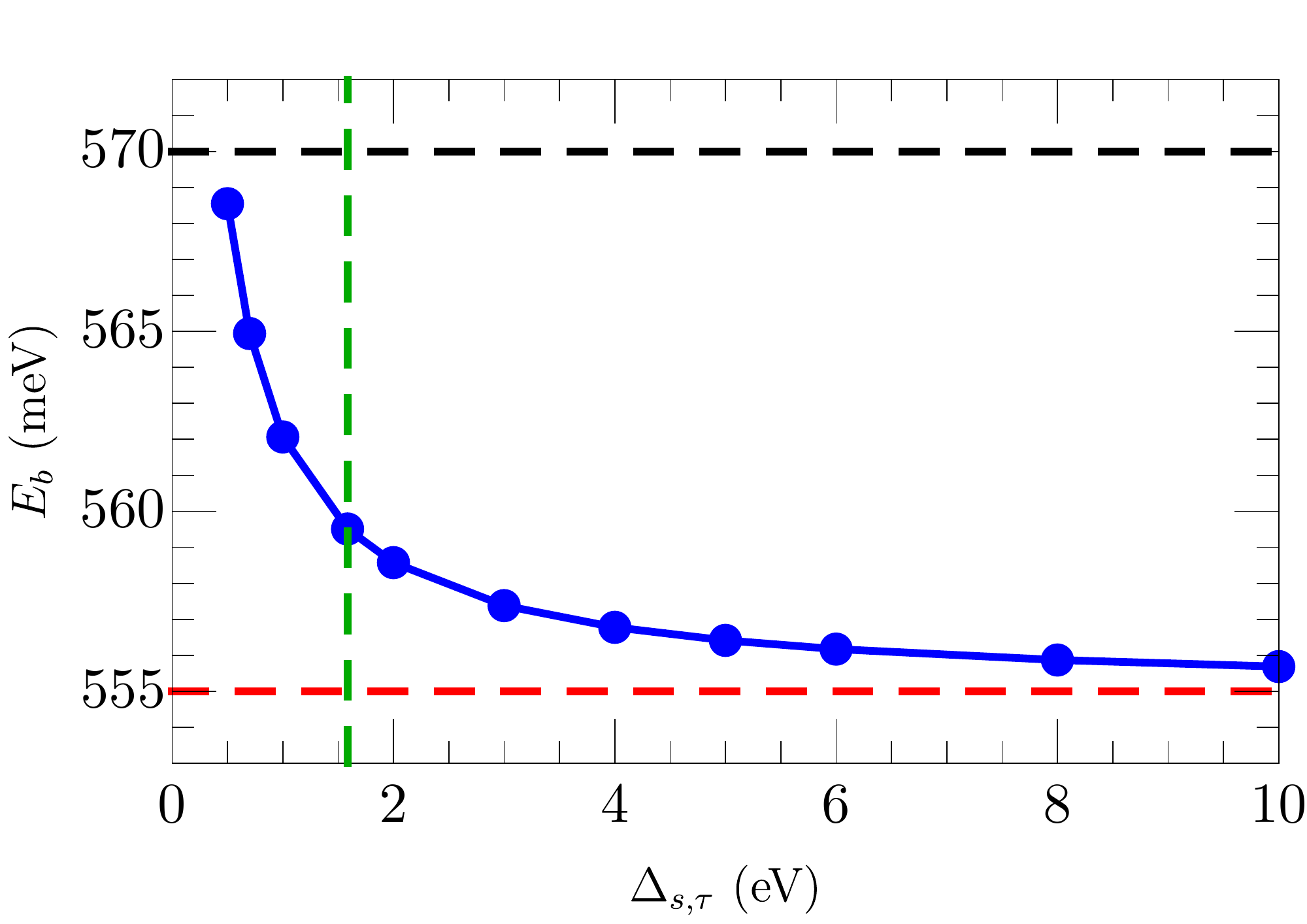}
\caption{(Color online) Binding energy for excitons in MoS$_2$ in vacuum as a function of the band gap for fixed charge carrier mass calculated in the multi-band model. The red and black dashed lines are the single-band SVM and experimental result, respectively. The green dashed line indicates the realistic value of the band gap of MoS$_2$.}
\label{fig:delta}
\end{figure}

In the single-band model, the difference in exciton energies as calculated with the FEM and SVM, respectively, is in the order of $10^{-3}$ meV. Therefore, in this subsection we only show SVM results for the single-band model.

In Fig. \ref{fig:delta} we show the binding energy as a function of the band gap. For each value of the band gap we calculate the value of $at$ such that it fixes the charge carrier mass at 0.5$m_0$ to facilitate comparison with the single-band model results and the experimental results. The figure shows that the binding energy calculated in the multi-band model converges to the binding energy calculated in the single-band model in the limit of an infinite band gap. As mentioned above, in the single-band model, the charge carrier mass replaces the role of the parameter $at$ in the multi-band model as the parameter that determines the curvature of the energy bands. However, the band gap parameter is effectively lost. Only conduction electrons (electrons in the conduction band) and conduction holes (the absence of electrons in the valence band) are considered in the single-band Hamiltonian \eqref{ham}, which can therefore be viewed as an infinite band gap approximation with the conduction band edge at zero energy and the valence band edge at minus infinity. As the band gap decreases, the binding energy increases. This is due to the interband interactions which become more important. It should be noted that when the band gap becomes of the same order as the binding energy, the latter will start to decrease since the exciton state will always be located inside the band gap. At the realistic value of the band gap of MoS$_2$, indicated by the green dashed line, we see that the multi-band model result agrees better with the experimental result than the single-band result.

\begin{figure}
\centering
\includegraphics[width=8.5cm]{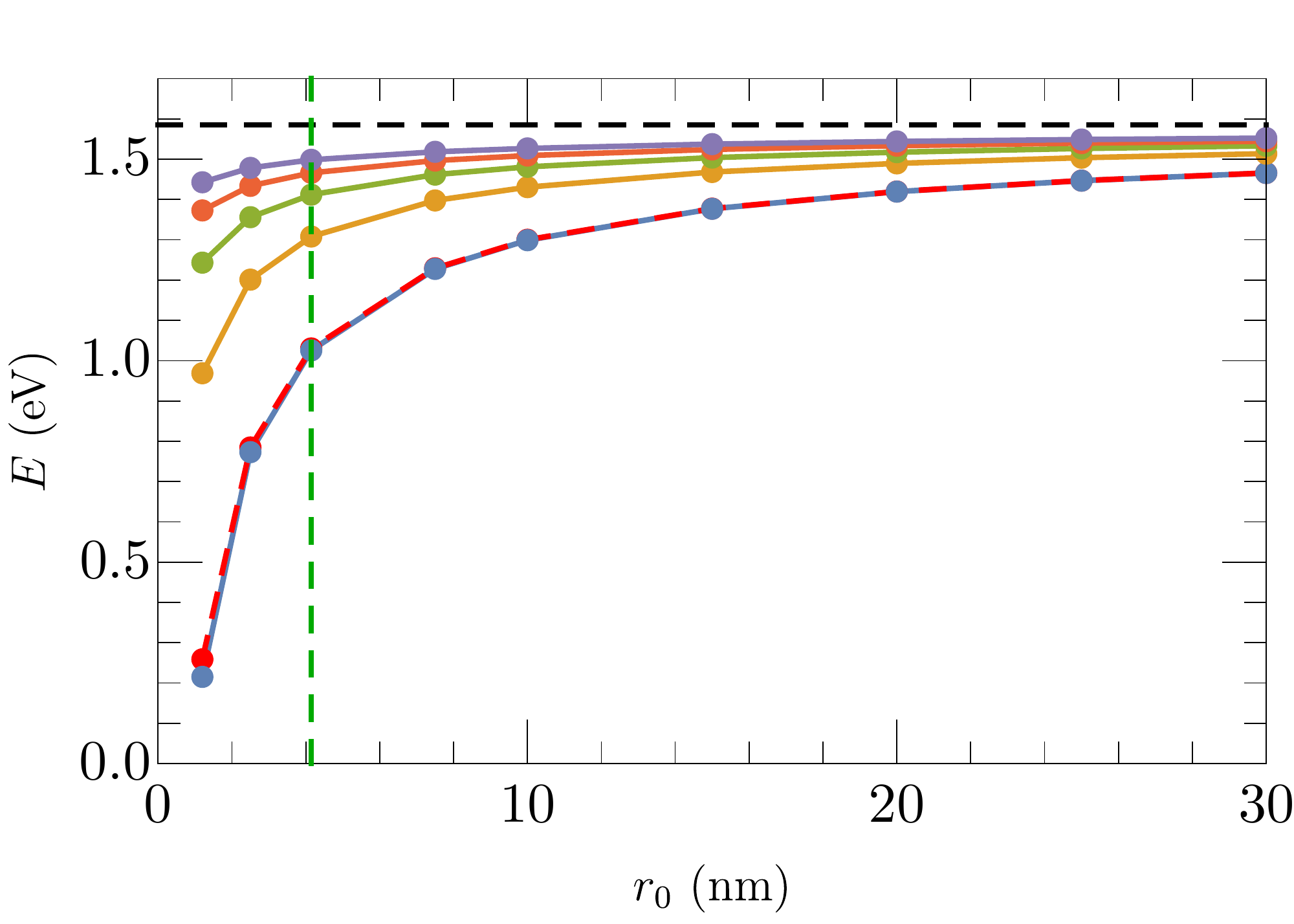}
\caption{(Color online) Five lowest energy levels for excitons in MoS$_2$ in vacuum as a function of the screening length calculated in the multi-band model. The red dashed curve is the single-band SVM result for the ground state. The black dashed line indicates the edge of the continuum. The green dashed line indicates the realistic value of the screening length of MoS$_2$.}
\label{fig:r0}
\end{figure}

The dependence of the five lowest energy levels of an exciton on the screening length is shown in Fig. \ref{fig:r0}. Similar to the case of the (2D) hydrogen atom, there are an infinite number of energy levels which pile up towards the edge of the continuum, which for the case of 2D TMDs lies at $\Delta_{s,\tau}$. As the screening length increases, and thus the interaction strength decreases, all the energy levels converge towards the edge of the continuum. In this limit the multi-band model results also agree perfectly with the single-band model results. For small screening lengths, however, the interactions are strong and therefore the contribution of the interband interactions becomes more important yielding significant differences between the multi-band and the single-band model. Note that, since the conduction band edge in the single-band model is located at zero energy, all the single-band model results are shifted by an amount $\Delta_{s,\tau}$ to facilitate comparison with the multi-band model results.

\begin{figure}
\centering
\includegraphics[width=8.5cm]{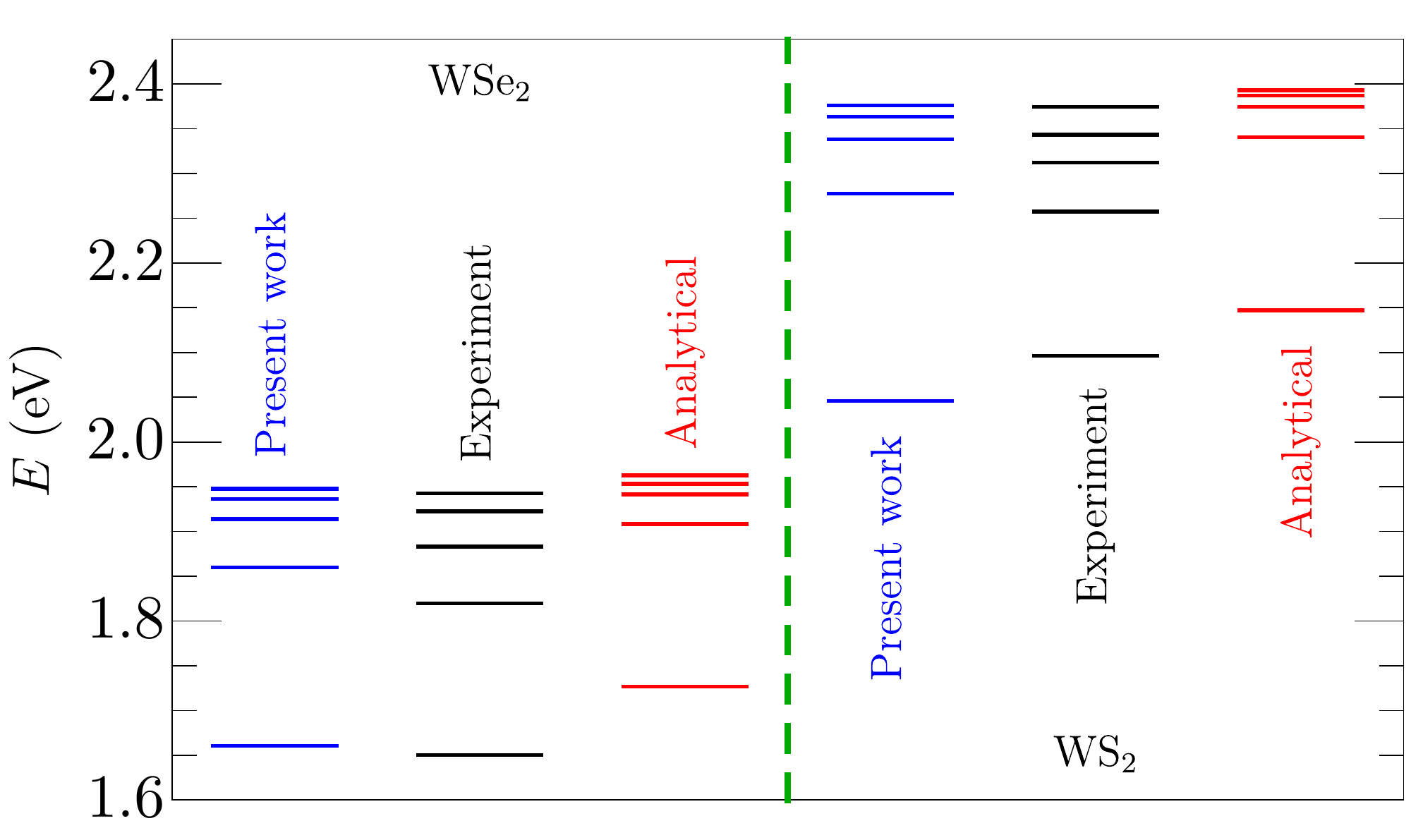}
\caption{(Color online) Five lowest energy levels for excitons in WS$_2$ (left panel) and WSe$_2$ (right panel) on a SiO$_2$ substrate, as determined from our FEM solution of the multi-band model (blue), the analytical model of Ref. [\onlinecite{excspecan}] (red), and the experimental results of Refs. [\onlinecite{chernikov}] (WS$_2$) and [\onlinecite{he}] (WSe$_2$) (black). For these calculations we used the parameters specified in Ref. [\onlinecite{excspecan}], i.e. $\Delta_{s,\tau}=2.4$ eV, $a=3.197$ \AA, and $t=1.25$ eV for WS$_2$, $\Delta_{s,\tau}=1.97$ eV, $a=3.310$ \AA, and $t=1.13$ eV for WSe$_2$, and a dielectric constant of $\varepsilon_r=3.9$ for the SiO$_2$ substrate.}
\label{fig:nlevel}
\end{figure}

In Fig. \ref{fig:nlevel} we show the five lowest energy levels for excitons in WS$_2$ and WSe$_2$ on a SiO$_2$ substrate calculated with the multi-band model and compare it with an analytical model\cite{excspecan} and experimental results\cite{chernikov,he}. This shows that the agreement of the multi-band model with the experimental results is better than that of the analytical model. Our results always overestimate the experimental results except for the ground state of WS$_2$ for which we find a smaller value.

\begin{figure}
\centering
\includegraphics[width=8.5cm]{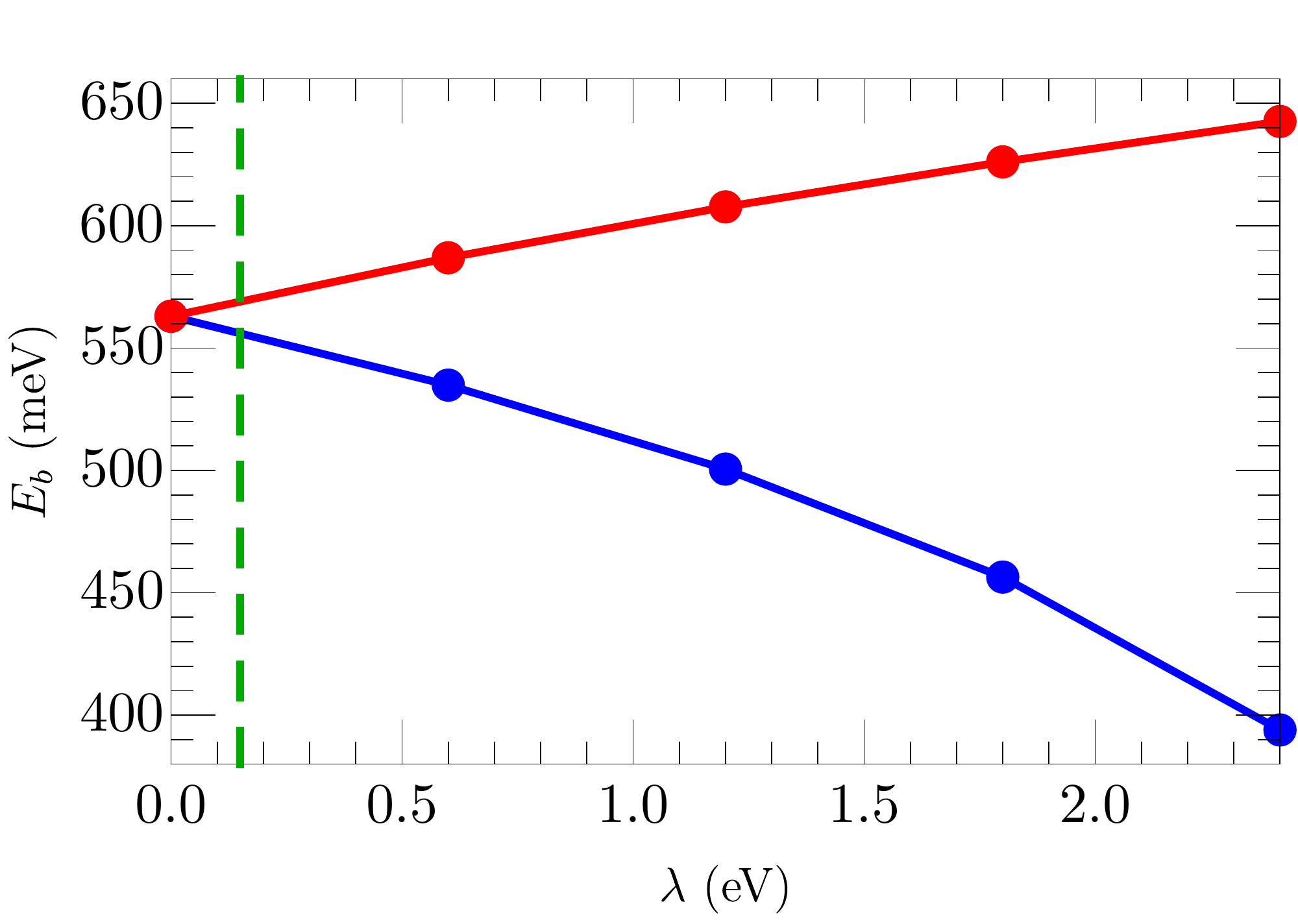}
\caption{(Color online) Binding energy for $A$ (blue) and $B$ (red) excitons in MoS$_2$ in vacuum as a function of the spin-orbit coupling strength calculated in the multi-band model. The green dashed line indicates the realistic value of the spin-orbit coupling strength of MoS$_2$.}
\label{fig:lambda}
\end{figure}

Due to the spin splitting of the valence band there are effectively two band gaps and as a consequence two different kinds of excitons. These are commonly referred to in the literature as $A$ and $B$ excitons and are illustrated in Fig. \ref{fig:schema}. When examining Eq. \eqref{difvgl}, we indeed see that there is a small difference in the equation depending on the value of $s^h\tau^h=\pm1$. When $s^h\tau^h=1$ Eq. \eqref{difvgl} describes the $A$ exciton and when $s^h\tau^h=-1$ it describes the $B$ exciton. To illustrate this, the $A$ excitons in Fig. \ref{fig:schema} in the $K$ and $K'$ valley have $s^e=1,\ \tau^e=1,\ s^h=-1,\ \tau^h=-1$ and $s^e=-1,\ \tau^e=-1,\ s^h=1,\ \tau^h=1$, respectively. The $B$ exciton in the figure has $s^e=-1,\ \tau^e=1,\ s^h=1,\ \tau^h=-1$. Note that the intervalley exciton, which can arise due to excitation of the charge carriers with linearly polarized light, also has $s^e\tau^e=s^h\tau^h$. In Fig. \ref{fig:lambda} we show the binding energy for $A$ and $B$ excitons as a function of the spin-orbit coupling strength. In the absence of spin-orbit coupling the binding energies for both kinds of excitons are equal since there is no spin splitting of the energy bands. For finite spin-orbit coupling, the $B$ exciton binding energy is always larger than the $A$ exciton binding energy, and this difference increases with increasing spin-orbit coupling since the $A$ exciton binding energy decreases whereas the $B$ exciton binding energy increases. Since increasing $\lambda$ will cause the $A$ exciton band gap to become smaller whereas the $B$ exciton band gap becomes larger, one may expect an increase (decrease) in the binding energy of the $A$ ($B$) exciton due to the respective changes in the band gap (see Fig. \ref{fig:delta}). However, while in Fig. \ref{fig:delta} we fixed the charge carrier mass, we now consider the more realistic case in which $at$ is fixed. By means of Eq. \eqref{mass} we see that an increasing (decreasing) band gap leads to an increasing (decreasing) charge carrier mass. This reduces (enhances) the kinetic energy and therefore enhances (reduces) the binding energy, thus explaining the results in the figure. At the realistic value of the spin-orbit coupling strength of MoS$_2$, indicated by the green dashed line, the difference in binding energy of the $A$ and $B$ exciton is 12.9 meV and the difference in the ground state energy of the $A$ and $B$ exciton is 137.1 meV. In the following of the present work we consider only the $A$ exciton.

\begin{figure}
\centering
\includegraphics[width=8.5cm]{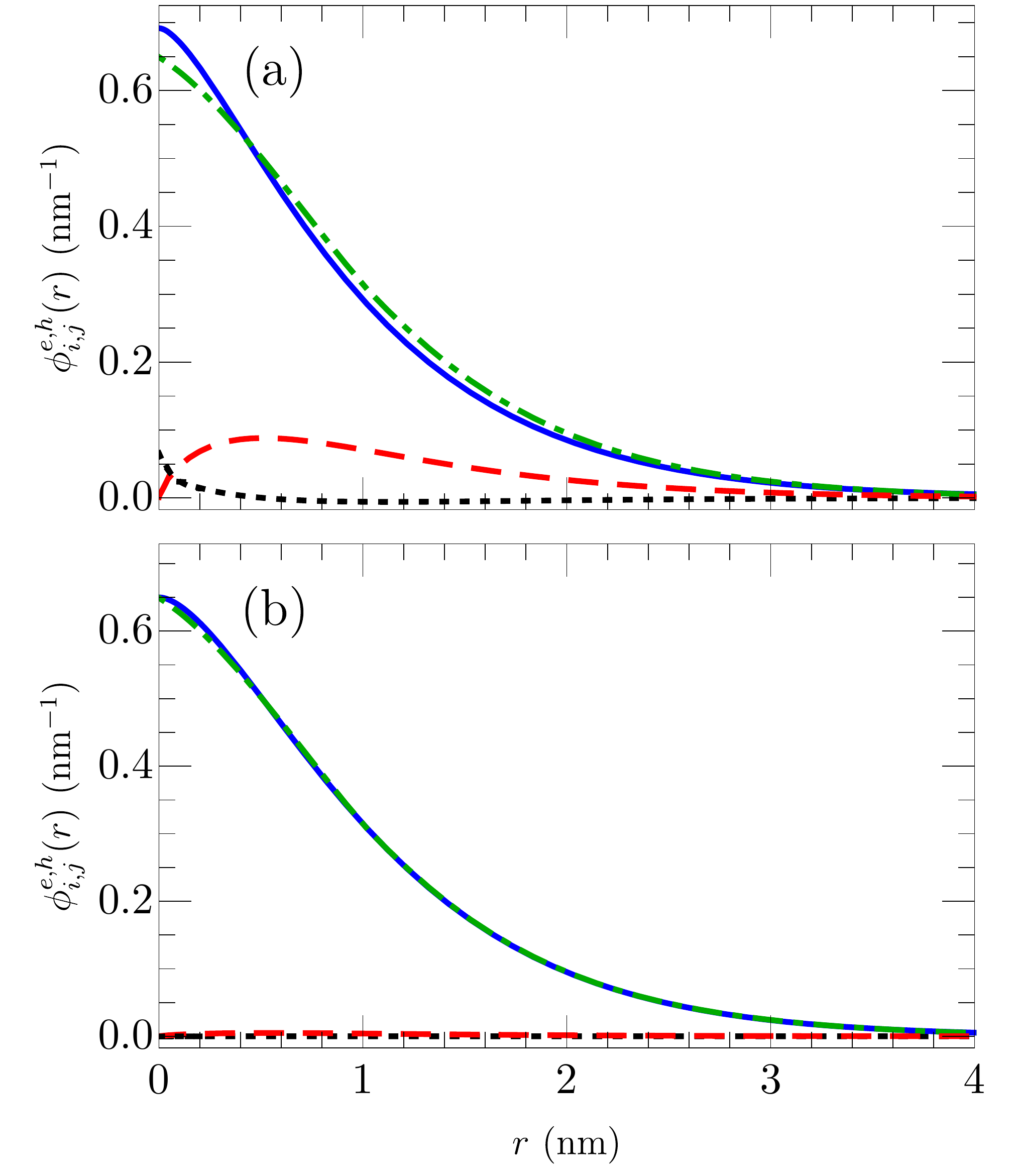}
\caption{(Color online) Different components ($i=c,j=v$: blue solid curve, $i=c$, $j=c$ and $i=v$, $j=v$: red dashed curve, $i=v$, $j=c$: black dotted curve) of the ground state wave function for excitons in MoS$_2$ in vacuum for $\Delta_{s,\tau}=1.585$ eV (a) and $\Delta_{s,\tau}=500$ eV (b) for fixed charge carrier mass calculated in the multi-band model. The green dot-dashed curve is the single-band SVM result.}
\label{fig:golf}
\end{figure}

In Fig. \ref{fig:golf} we show the different components of the exciton ground state wave function. For the realistic value of the band gap, we see that the component with the electron in the conduction band and the hole in the valence band is the most important one, it is an order of magnitude larger than the component with both particles in the conduction band and the component with both particles in the valence band (these two components are identical) and almost two orders of magnitude larger than the component with the electron in the valence band and the hole in the conduction band. Furthermore, the single-band model wave function is in good agreement with the dominant component. Note that the single-band model wave function only has one component since in this model only conduction particles are considered. It is precisely the other components, representing the contribution of excitons consisting partly or entirely of valence particles, which lead to the increasing binding energy with decreasing band gap shown in Fig. \ref{fig:delta}. For the case of a very large band gap, we see that these valence components are completely suppressed and that the dominant conduction component agrees perfectly with the single-band model wave function, which is in correspondence with the fact that the binding energy in the multi-band model converges to the single-band model binding energy in the limit of an infinite band gap.

\begin{figure}
\centering
\includegraphics[width=8.5cm]{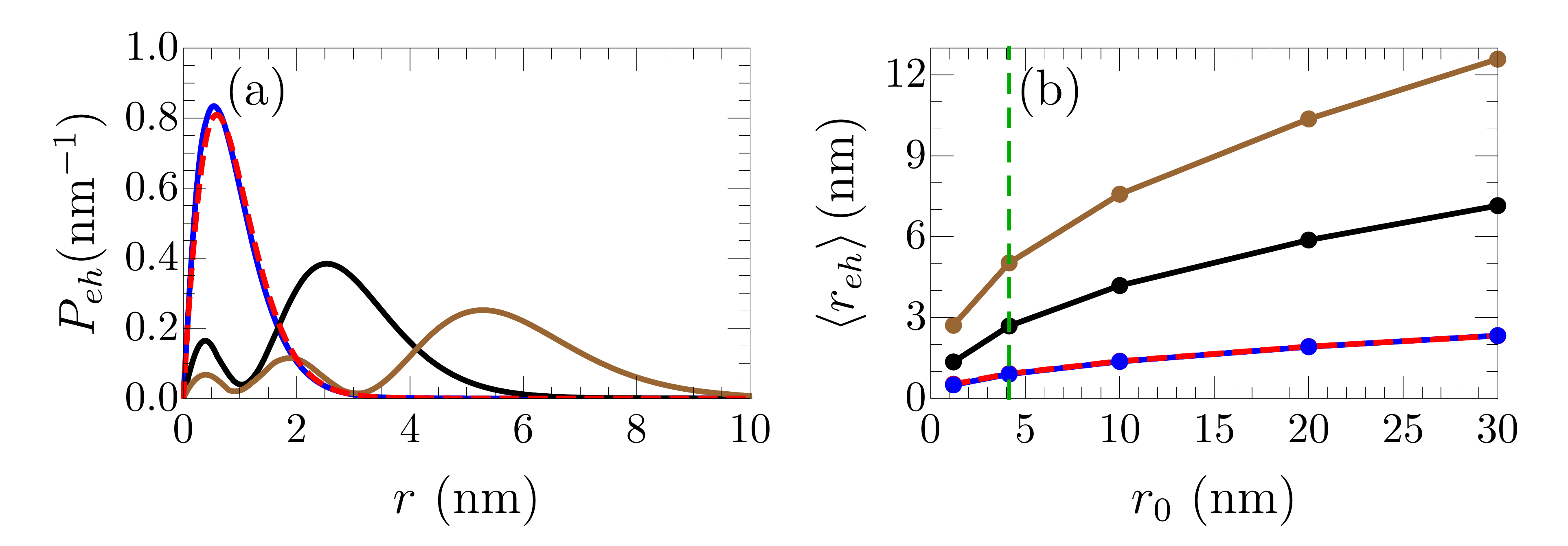}
\caption{(Color online) Interparticle distance probability distribution (a) and average interparticle distance as a function of the screening length (b) for the first three states for excitons in MoS$_2$ in vacuum calculated in the multi-band model. The red dashed curve is the single-band SVM result for the ground state. The green dashed line indicates the realistic value of the screening length of MoS$_2$.}
\label{fig:cd}
\end{figure}

The interparticle distance probability distributions for the first three states of an exciton are shown in Fig. \ref{fig:cd}(a). This shows that the excited states have a larger probability for the particles being at a larger distance from each other. In general, the interparticle distance probability distribution of the $n^{\text{th}}$ state exhibits $n$ maxima. Furthermore, the single-band model result agrees very well with the multi-band model result. In Fig. \ref{fig:cd}(b) we show the average interparticle distance for the first three states as a function of the screening length. This shows that the average interparticle distance increases with the screening length, which is a consequence of the reduced interaction strength, as well as with increasingly excited states. Again, the single-band model result agrees well with the multi-band model result.

\begin{figure}
\centering
\includegraphics[width=8.5cm]{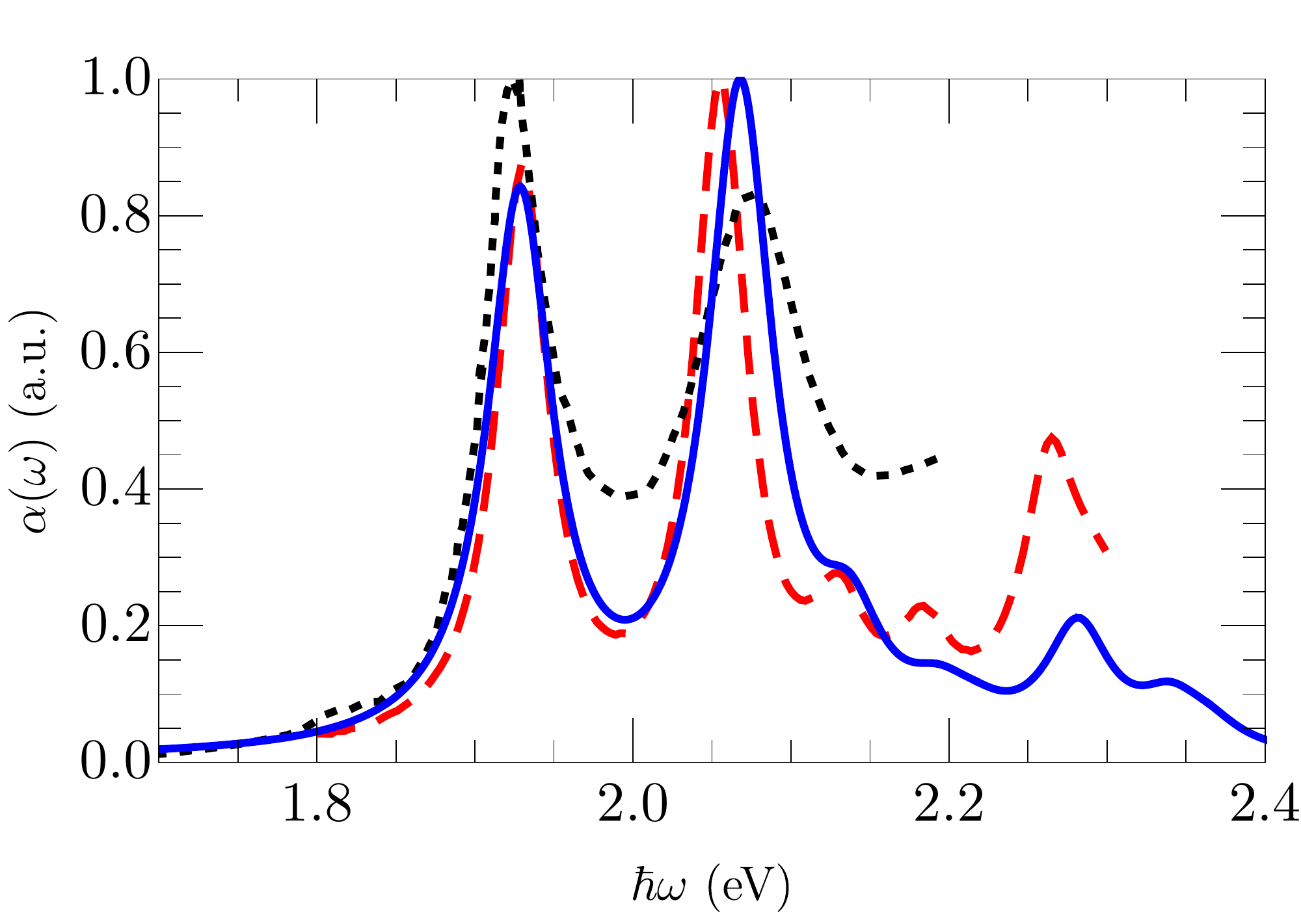}
\caption{(Color online) Excitonic absorbance spectrum for MoS$_2$ on a SiO$_2$ substrate, as determined from our FEM solution of the multi-band model (blue solid curve), the numerical solution of the Bethe-Salpeter equation of Ref. [\onlinecite{absorbthe}] (red dashed curve), and the experimental results of Ref. [\onlinecite{mak3}] (black dotted curve). Note that the results of the two theoretical results are shifted such that they match the $A$ exciton energy of the experimental results. The results of the different models are also rescaled to facilitate comparison. For our numerical results we used a broadening of 25 meV.}
\label{fig:absorptie}
\end{figure}

We show the optical absorption spectrum in Fig. \ref{fig:absorptie}. Our results obtained with the multi-band model are in good agreement with results obtained from numerically solving the Bethe-Salpeter (BS) equation\cite{absorbthe}. The most noticeable difference between the two is the height of the peak around 2.27 eV, which corresponds to the $2s$-state of the $B$ exciton, which is about 50\% of the peak of the $1s$-state of the $B$ exciton in the results obtained from the BS equation whereas it is about 20\% of this peak in our multi-band model. The splitting between the peaks of the $1s$-state of the $A$ and the $B$ exciton is similar for the three different results shown in this figure. However, both the multi-band model and the BS equation predict that the $B$ exciton peak is higher than the $A$ exciton peak whereas in the experimental results the $A$ exciton peak is more pronounced than the $B$ exciton peak. It should be noted that the experimental absorbance was not measured up to high enough photon energies to clearly distinguish the excited states.

\begin{table}
\centering
\caption{Exciton binding energies (meV) for different TMD materials in the single-band model obtained with the SVM and in the multi-band model (MB), compared with previous theoretical results in the single-band model and experimental studies. Results of the present work are listed in bold. We used $\varepsilon_r=3.8$ for SiO$_2$ and $\varepsilon_r=4.58$ for bilayer graphene (BLG).}
\begin{tabular}{ccccccc}
\hline
\hline
 & Substrate & Theory & Experiment & SVM & MB \\
\hline
\hline
Mo$\text{S}_2$ & Vacuum & 551.4 [\onlinecite{theory}] & 570 [\onlinecite{MoS2exc2}] & 555.0 [\onlinecite{analytic}] & \textbf{559.5} \\
 & & 526.5 [\onlinecite{theory2}] & & & \\
 & Si$\text{O}_2$ & 348.6 [\onlinecite{theory2}] & & \textbf{320.4} & \textbf{323.9} \\
\hline
MoS$\text{e}_2$ & Vacuum & 477.8 [\onlinecite{theory}] & & 480.4 [\onlinecite{analytic}] & \textbf{483.8} \\
 & & 476.9 [\onlinecite{theory2}] & & & \\
 & Si$\text{O}_2$ & 322.9 [\onlinecite{theory2}] & & \textbf{286.1} & \textbf{288.8} \\
 & BLG & & 550 [\onlinecite{MoSe2exc}] & \textbf{256.7} & \textbf{259.2} \\
 & & & 580 [\onlinecite{MoSe2exc2}] & & \\
\hline
W$\text{S}_2$ & Vacuum & 519.1 [\onlinecite{theory}] & & 523.5 [\onlinecite{analytic}] & \textbf{528.6} \\
 & & 509.8 [\onlinecite{theory2}] & & & \\
 & Si$\text{O}_2$ & 322.9 [\onlinecite{theory2}] & 320 [\onlinecite{he}] & \textbf{284.1} & \textbf{287.5} \\
 & & & 312 [\onlinecite{perpfield}] & & \\
\hline
WS$\text{e}_2$ & Vacuum & 466.7 [\onlinecite{theory}] & & 470.2 [\onlinecite{analytic}] & \textbf{474.4} \\
 & & 456.4 [\onlinecite{theory2}] & & \\
 & Si$\text{O}_2$ & 294.6 [\onlinecite{theory2}] & 370 [\onlinecite{he}] & \textbf{262.5} & \textbf{265.5} \\
\hline
\hline
\end{tabular}
\label{table:exctable}
\end{table}

In Table \ref{table:exctable} we show the exciton binding energies for different TMD materials and substrates in the single-band model obtained with the SVM and in the multi-band model, compared with theoretical studies in the single-band model using ground-state diffusion Monte Carlo\cite{theory} and path-integral Monte Carlo\cite{theory2}, as well as experimental results. The results obtained in the multi-band model are in good agreement with the other theoretical and SVM results, which are all calculated in the single-band model. Note that the authors of Ref. [\onlinecite{theory2}] use a value of $\varepsilon_r=3.0$ for SiO$_2$, which explains why their binding energies in this case are consistently larger than ours. If we take $\varepsilon_r=3.0$ we find a binding energy of 365.2 meV, 323.5 meV, 328.6 meV, and 301.5 meV for MoS$_2$, MoSe$_2$, WS$_2$, and WSe$_2$, respectively, with the SVM. This agrees well with the results of Ref. [\onlinecite{theory2}]. We find that the multi-band model results are larger by about 3-5 meV as compared to the single-band SVM results and as such are closer to the experimental results. The agreement with the experimental results is good for MoS$_2$ and for WS$_2$, whereas for WSe$_2$ and especially for MoSe$_2$ the agreement with the experimental results is less good. Although, for the latter, a possible explanation is given by the fact that the use of bilayer graphene, a strictly two-dimensional material, as a substrate must be taken into account in a different way as compared to just inserting its dielectric constant in the screened interaction potential \eqref{interpot}. However, it is also remarkable that the experimental result in the presence of a substrate is about 100 meV larger than the theoretically predicted result in vacuum.

\subsection{Trion}

\begin{figure}
\centering
\includegraphics[width=8.5cm]{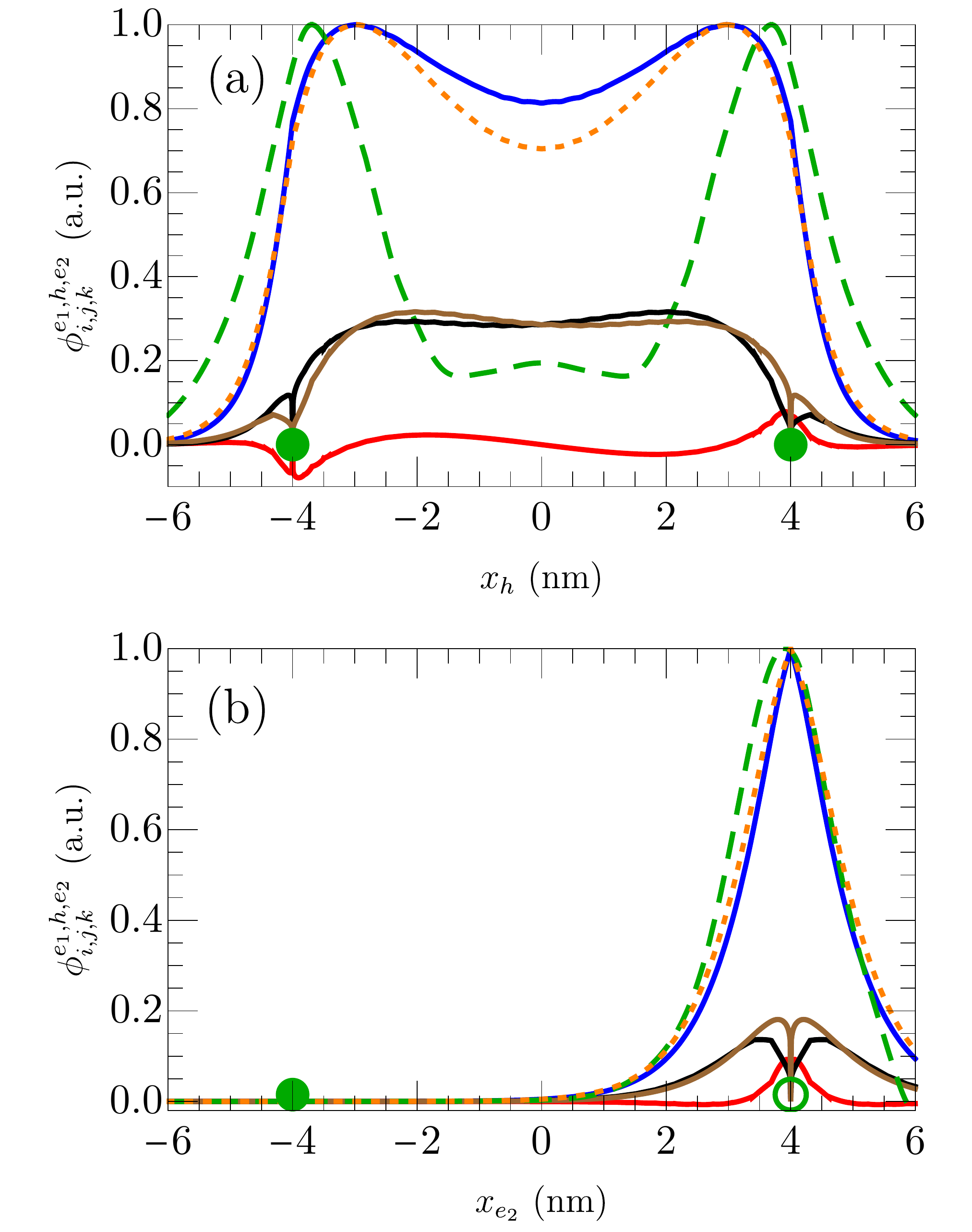}
\caption{(Color online) (a) Different components ($i=c,j=v,k=c$: blue curve, $i=c$, $j=c,k=c$: red curve $i=c$, $j=v,k=v$: black curve, $i=v$, $j=v,k=c$: brown curve) of the ground state wave function for trions in MoS$_2$ in vacuum as a function of the hole coordinate $x_h$ for $y_h=0$ and for fixed electrons calculated in the multi-band model. The green dashed curve and orange dotted curve are the single-band SVM and FEM result, respectively. (b) Same as (a) but now as a function of the electron coordinate $x_{e_2}$ for $y_{e_2}=0$ and for a fixed electron and hole. Open and closed circles indicate the position of holes and electrons, respectively.}
\label{fig:golftri}
\end{figure}

Here we limit ourselves to trions consisting of an $A$ exciton and an additional electron. In Fig. \ref{fig:golftri} we show the different components of the trion ground state wave function. Similar to what we found for the case of the exciton, we now have one dominant component which represents a trion consisting of three conduction particles, three components which are an order of magnitude smaller representing a trion consisting of two conduction particles and one valence particle, three components which are two orders of magnitude smaller representing a trion consisting of one conduction particle and two valence particles and one component which is three orders of magnitude smaller representing a trion consisting of three valence particles. In Fig. \ref{fig:golftri}, we only show the four largest components.

We show the wave function as a function of the hole $x$-coordinate when its $y$-coordinate and the two electrons are fixed in Fig. \ref{fig:golftri}(a). This shows that the hole localizes equally around the two electrons. The component $\phi^{e_1,h,e_2}_{c,c,c}$ of the wave function also shows extrema around the electron positions whereas the main contribution of the other two non-dominant components is in between the two electrons. Furthermore, the SVM wave function shows qualitatively the same behavior as the dominant multi-band model wave function component and the single-band FEM wave function but there are substantial quantitative differences. These differences are a consequence of the fact that we neglected the dependence of the wave function on the angular coordinates when using the FEM.

In Fig. \ref{fig:golftri}(b) we show the wave function as a function of the electron $x$-coordinate when its $y$-coordinate and the hole and the other electron are fixed, which shows that the electron localizes around the hole. The other components show a similar behavior as the dominant component. In this case the agreement between the SVM and FEM wave functions is better than in the case of fixed electrons. 

\begin{figure}
\centering
\includegraphics[width=8.5cm]{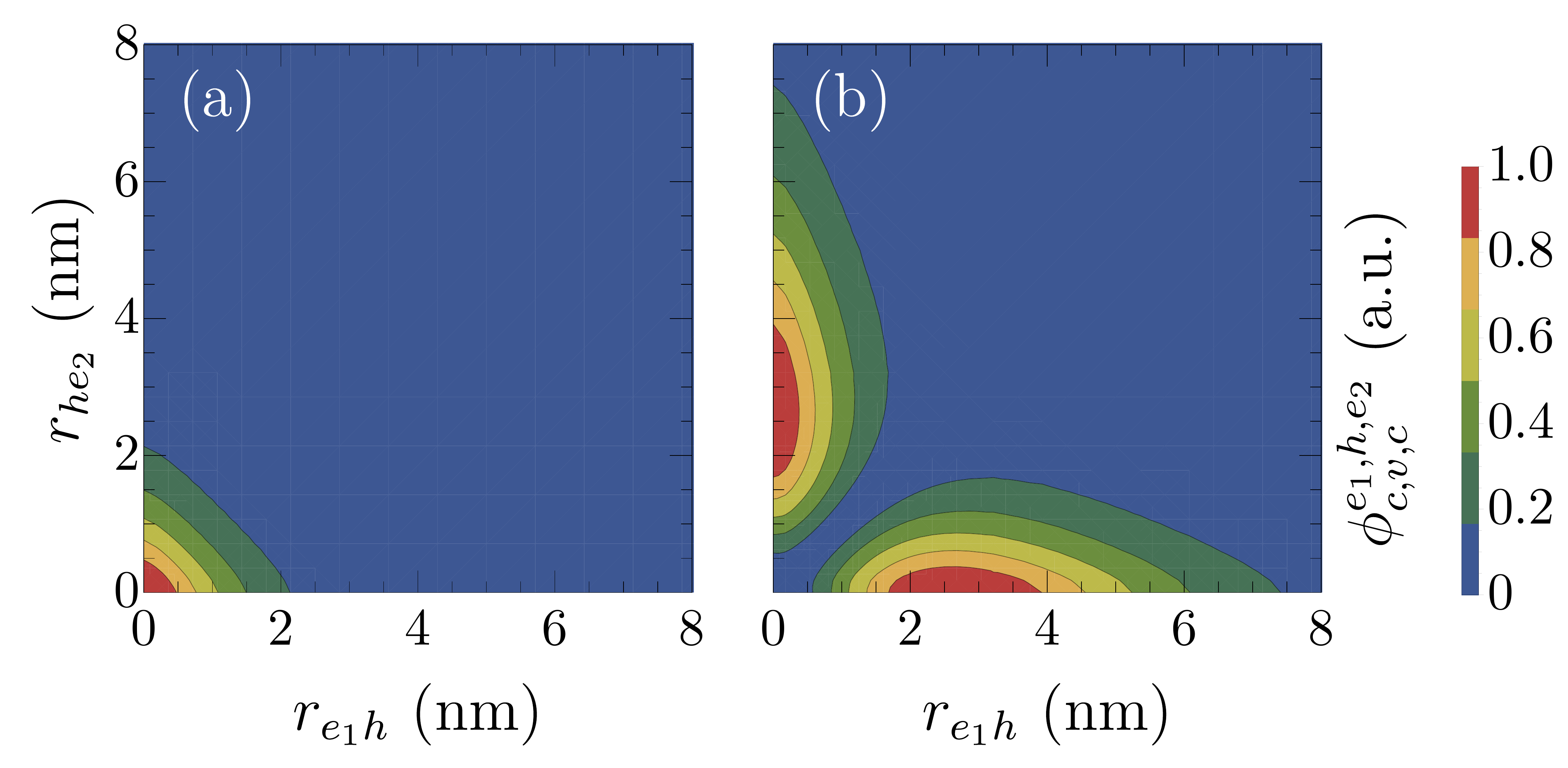}
\caption{(Color online) Dominant component $\phi^{e_1,h,e_2}_{c,v,c}(r_{e_1h},r_{he_2})$ of the ground state (a) and second excited state (b) wave function for trions in MoS$_2$ in vacuum as a function of the relative coordinates calculated in the multi-band model.}
\label{fig:golftric}
\end{figure}

\begin{table}
\centering
\caption{Negative trion binding energies (meV) for different TMD materials in the single-band model obtained with both the SVM and the FEM and in the multi-band model (MB), compared with previous theoretical results in the single-band model and experimental studies. Results of the present work are listed in bold. We used $\varepsilon_r=3.8$ for SiO$_2$.}
\begin{tabular}{cccccccc}
\hline
\hline
 & Substrate & Theory & Exp. & SVM & FEM & MB \\
\hline
\hline
Mo$\text{S}_2$ & Vacuum & 33.8 [\onlinecite{theory}] & & 33.7 [\onlinecite{analytic}] & \textbf{23.1} & \textbf{37.9} \\
 & & 32.0 [\onlinecite{theory2}] & & & \\
 & & 32 [\onlinecite{falkomc}] & & & \\
 & Si$\text{O}_2$ & 24.7 [\onlinecite{theory2}] & 18 [\onlinecite{mak3}] & \textbf{23.2} & \textbf{3.4} & \textbf{14.5} & \\
\hline
MoS$\text{e}_2$ & Vacuum & 28.4 [\onlinecite{theory}] & & 28.2 [\onlinecite{analytic}] & \textbf{21.1} & \textbf{32.5} \\
 & & 27.7 [\onlinecite{theory2}] & & & \\
 & & 31 [\onlinecite{falkomc}] & & & \\
  & Si$\text{O}_2$ & 22.1 [\onlinecite{theory2}] & 30 [\onlinecite{expT3}] & \textbf{19.7} & \textbf{5.3} & \textbf{14.2} \\
\hline
W$\text{S}_2$ & Vacuum & 34.0 [\onlinecite{theory}] & & 33.8 [\onlinecite{analytic}] & \textbf{18.8} & \textbf{35.2} \\
 & & 33.1 [\onlinecite{theory2}] & & & \\
 & & 31 [\onlinecite{falkomc}] & & & \\
 & Si$\text{O}_2$ & 24.3 [\onlinecite{theory2}] & 30 [\onlinecite{WS2tri}] & \textbf{20.2} & \textbf{-1.9} & \textbf{9.4} \\
 & & & 30 [\onlinecite{WS2tri2}] & & & \\
\hline
WS$\text{e}_2$ & Vacuum & 29.5 [\onlinecite{theory}] & & 29.5 [\onlinecite{analytic}] & \textbf{18.3} & \textbf{32.1} \\
 & & 28.5 [\onlinecite{theory2}] & & & \\
 & & 27 [\onlinecite{falkomc}] & & & \\
 & Si$\text{O}_2$ & 21.5 [\onlinecite{theory2}] & 30 [\onlinecite{WSe2tri}] & \textbf{19.3} & \textbf{0.4} & \textbf{10.4} \\
\hline
\hline
\end{tabular}
\label{table:tritable}
\end{table}

The dominant component of the ground state wave function and the second excited state wave function is shown as a function of the relative coordinates in Fig. \ref{fig:golftric}. The ground state wave function has a single maximum at $r_{e_1h}=r_{he_2}=0$ meaning that the three particles form one symmetric system with comparable average interparticle distances between all the particles. We find $\braket{r_{eh}}\approx0.96$ nm and $\braket{r_{ee}}\approx1.40$ nm, yielding a ratio of $\braket{r_{eh}}/\braket{r_{ee}}=0.69$. The second excited state however has two maxima: one at $r_{e_1h}=0$ and $r_{he_2}\approx3$ nm and one at $r_{e_1h}\approx3$ nm and $r_{he_2}=0$. This means that the structure of the trion is now given by an exciton, consisting of one of the electrons and the hole, with the additional electron circling around it. In this case we find $\braket{r_{eh}}\approx2.31$ nm and $\braket{r_{ee}}\approx3.81$ nm. However, since the two electrons are identical, this implies that $\braket{r_{eh}}$ is the average of the average distance between the hole and the inner electron and the average distance between the hole and the outer electron. Approximating the latter by the average electron-electron distance we find that the average distance between the hole and the inner electron is given by $\braket{r_{eh}^{in}}\approx0.80$ nm. This gives $\braket{r_{eh}^{in}}/\braket{r_{ee}}=0.21$.

In Table \ref{table:tritable} we show the trion binding energies for different TMD materials and substrates in the single-band model obtained with both the SVM and the FEM and in the multi-band model, compared with theoretical studies in the single-band model using ground-state diffusion Monte Carlo\cite{theory,falkomc} and path-integral Monte Carlo\cite{theory2}, as well as experimental results. For TMDs suspended in vacuum the results obtained with the multi-band model are larger than the SVM results by a similar amount as in the case of excitons. For TMDs placed on a SiO$_2$ substrate the multi-band model results are lower than the SVM results. Again note that $\varepsilon_r=3.0$ is used for SiO$_2$ in Ref. [\onlinecite{theory2}], explaining their larger binding energies in this case. If we take $\varepsilon_r=3.0$ we find a binding energy of 24.9 meV, 21.6 meV, 24.0 meV, and 20.5 meV for MoS$_2$, MoSe$_2$, WS$_2$, and WSe$_2$, respectively, with the SVM. This agrees well with the results of Ref. [\onlinecite{theory2}]. Furthermore, the single-band FEM results agree very badly with the single-band SVM results, even predicting unstable trions in WS$_2$, showing the importance of angular correlations in trions. In general, the agreement with the experimental results is not as good as for the case of excitons. In a recent experiment\cite{nieuw} a binding energy of 21 (32) meV was found for positive (negative) trions in WSe$_2$ encapsulated in hexagonal boron nitride layers. This difference was attributed to a combination of a small difference in the effective mass of the two charge carriers and exchange interaction effects.

\section{Summary and conclusion}
\label{sec:Summary and conclusion}

In this paper, we studied the electronic and structural properties of excitons and trions in 2D transition metal dichalcogenides. We considered a multi-band model taking into account the full low-energy dispersion for monolayer TMDs including the spin-orbit coupling and solved it using the finite element method. We also considered a simplified single-band model which we solved with both the finite element method as well as the stochastic variational method.

Calculating the excitonic energies and wave functions in the single-band model with the FEM instead of the SVM has the advantages that it allows to readily obtain the excited excitonic states and that the calculations are computationally about a factor 5 faster than the SVM calculations. We found nearly perfect agreement between the results of both methods and good agreement with other theoretical results for the case of excitons. For the case of trions, we found significant differences between the two methods. This is due to the neglect of the angular dependence of the wave function when using the FEM. Therefore, we can conclude that angular correlations are important in trions and as a result that the SVM is needed to obtain good quantitative results. Furthermore, for biexcitons and even larger excitonic systems, calculating the excitonic energies and wave functions will become almost impossible with the FEM while it is still feasible with the SVM.

The agreement of these theoretical models with experimental results is reasonable, although there are significant differences. A possible explanation can be that, considering the very small size of the excitonic systems, i.e. of the order of a few lattice constants, a continuum approach may lead to considerable errors and a discrete lattice approach may be necessary to obtain very good quantitative agreement with experiments.

Furthermore, we found that the exciton binding energy in the multi-band model converges towards the single-band model result in the limit of an infinite band gap. For finite band gaps the multi-band model result is larger than the single-band model result. We were able to explain this by means of the contribution of interband interactions. The multi-band model therefore allows to take into account the contribution of excitonic systems consisting partly or entirely out of valence particles. This was confirmed by plots of the different components of the wave functions of the excitonic systems. Another advantage of the multi-band model is that it allows for a more straightforward analysis of $A$ and $B$ excitons. Finally, we can conclude that it should be interesting to determine the energies and wave functions for trions and biexcitons in the multi-band model using the SVM.

\section{Acknowledgments}

This work was supported by the Research Foundation of Flanders (FWO-Vl) through an aspirant research grant for MVDD.

\appendix

\begin{widetext}
\section{Decoupling of the exciton eigenvalue equation}
\label{sec:appA}

Due to the presence of the $V(|\vec{r}_e-\vec{r}_h|)I_4$ term, the Hamiltonian does not commute with $\vec{k}^e$ nor with $\vec{k}^h$. This means that the components of the single-particle wave vectors are not good quantum numbers and should be replaced by their corresponding differential operators when solving the eigenvalue problem in the position representation. However, if we transform the single-particle coordinates to center of mass and relative coordinates,
\begin{equation}
\label{cortransf}
\vec{R} = \frac{\vec{r}_e+\vec{r}_h}{2}, \quad \vec{r} = \vec{r}_e-\vec{r}_h, \quad \vec{K} = \vec{k}^e+\vec{k}^h, \quad \vec{k} = \frac{\vec{k}^e-\vec{k}^h}{2},
\end{equation}
the interaction term becomes $V(r)I_4$. As a consequence, the Hamiltonian does not commute with the relative wave vector $\vec{k}$ but does commute with the center of mass momentum $\vec{K}$. Therefore, $\vec{K}$ is a conserved quantity and its components are good quantum numbers. Since we are only interested in the exciton states with the lowest energy, we take $\vec{K}=\vec{0}$ to discard the translational kinetic energy. Defining
\begin{equation}
\label{Odef}
\mathcal{O}_e = at(\tau^ek_x-ik_y)I_2, \quad \mathcal{O}_h = -at(-\tau^hk_x\sigma_x+k_y\sigma_y),
\end{equation}
the exciton eigenvalue equation \eqref{eigen} can be rewritten as
\begin{equation}
\label{eigen2}
\begin{cases}
\left(\mathcal{O}_h-V(r)I_2+\frac{\Delta-\lambda s^h\tau^h}{2}(I_2-\sigma_z)\right)\ket{\Psi^e_c}+\mathcal{O}_e\ket{\Psi^e_v} = E^{exc}_{\alpha}\ket{\Psi^e_c} \\
\mathcal{O}_e^{\dag}\ket{\Psi^e_c}+\left(\mathcal{O}_h-V(r)I_2-\frac{\Delta-\lambda s^e\tau^e}{2}(I_2+\sigma_z)+\frac{\lambda(s^e\tau^e-s^h\tau^h)}{2}(I_2-\sigma_z)\right)\ket{\Psi^e_v} = E^{exc}_{\alpha}\ket{\Psi^e_v}
\end{cases},
\end{equation}
with $\ket{\Psi^e_c}=\left(\ket{\phi^{e,h}_{c,c}},\ket{\phi^{e,h}_{c,v}}\right)^T$ and $\ket{\Psi^e_v}=\left(\ket{\phi^{e,h}_{v,c}},\ket{\phi^{e,h}_{v,v}}\right)^T$. It follows from the second equation that
\begin{equation}
\label{rel1}
\ket{\Psi^e_v} \approx \left(E^{exc}_{\alpha}I_2+V(r)I_2+\frac{\Delta-\lambda s^e\tau^e}{2}(I_2+\sigma_z)-\frac{\lambda(s^e\tau^e-s^h\tau^h)}{2}(I_2-\sigma_z)\right)^{-1}\mathcal{O}_e^{\dag}\ket{\Psi^e_c},
\end{equation}
where we have assumed the relative kinetic energy to be small compared to the band gap and the exciton energy. Using this result, the first equation of \eqref{eigen2} can be written as
\begin{equation}
\label{eigen3}
\begin{cases}
\left(-V(r)+\frac{a^2t^2k^2}{E^{exc}_{\alpha}+V(r)+\Delta-\lambda s^e\tau^e}+a^2t^2\left((\tau^ek_x-ik_y)\frac{1}{E^{exc}_{\alpha}+V(r)+\Delta-\lambda s^e\tau^e}\right)(\tau^ek_x+ik_y)\right)\ket{\phi^{e,h}_{c,c}} \\
\hspace{30pt}+at(\tau^hk_x+ik_y)\ket{\phi^{e,h}_{c,v}} = E^{exc}_{\alpha}\ket{\phi^{e,h}_{c,c}} \\
\hspace{6pt}at(\tau^hk_x-ik_y)\ket{\phi^{e,h}_{c,c}}+\Big(-V(r)+\Delta-\lambda s^h\tau^h \\
\hspace{30pt}+\frac{a^2t^2k^2}{E^{exc}_{\alpha}+V(r)-\lambda(s^e\tau^e-s^h\tau^h)}+a^2t^2\left((\tau^ek_x-ik_y)\frac{1}{E^{exc}_{\alpha}+V(r)-\lambda(s^e\tau^e-s^h\tau^h)}\right)(\tau^ek_x+ik_y)\Big)\ket{\phi^{e,h}_{c,v}} = E^{exc}_{\alpha}\ket{\phi^{e,h}_{c,v}}
\end{cases}.
\end{equation}
From the first equation we now have
\begin{equation}
\label{rel2}
\ket{\phi^{e,h}_{c,c}} \approx \frac{at(\tau^hk_x+ik_y)}{E^{exc}_{\alpha}+V(r)}\ket{\phi^{e,h}_{c,v}},
\end{equation}
again assuming the relative kinetic energy to be small compared to the exciton energy. Inserting this in the second equation of \eqref{eigen3} and going to position representation we get
\begin{equation}
\label{eigen4}
\begin{split}
\bigg(&-a^2t^2\left(\frac{1}{E^{exc}_{\alpha}+V(r)}+\frac{1}{E^{exc}_{\alpha}+V(r)-\lambda(s^e\tau^e-s^h\tau^h)}\right)\nabla^2_{\vec{r}}-V(r)+\Delta-\lambda s^h\tau^h \\
&-a^2t^2\left(\frac{\partial}{\partial r}\frac{1}{E^{exc}_{\alpha}+V(r)}+\frac{\partial}{\partial r}\frac{1}{E^{exc}_{\alpha}+V(r)-\lambda(s^e\tau^e-s^h\tau^h)}\right)\frac{\partial}{\partial r}\bigg)\phi^{e,h}_{c,v}(r) = E^{exc}_{\alpha}\phi^{e,h}_{c,v}(r),
\end{split}
\end{equation}
where we have used
\begin{equation}
\label{exterm}
\left[(\tau k_x-ik_y)f(r)\right](\tau k_x+ik_y)\phi(r) = -\left[\left(\tau e^{-i\tau\varphi}\partial_r-\frac{i}{r}e^{-i\tau\varphi}\partial_{\varphi}\right)f(r)\right]\left(\tau e^{i\tau\varphi}\partial_r+\frac{i}{r}e^{i\tau\varphi}\partial_{\varphi}\right)\phi(r) = -\left[\partial_rf(r)\right]\partial_r\phi(r),
\end{equation}
where $f(r)$ is a general function which only depends on the radial coordinate. For $s$-states the wave function $\phi(r)$ does not depend on the angular coordinate. The energy levels and the component $\phi^{e,h}_{c,v}(r)$ of the wave function can be determined from Eq. \eqref{eigen4}. The other three components of the wave function can be determined from Eqs. \eqref{rel1} and \eqref{rel2}. For $s^e\tau^e=s^h\tau^h$ Eq. \eqref{eigen4} reduces to Eq. \eqref{difvgl}.

The trion eigenvalue problem in Eq. \eqref{eigtri} can be decoupled in a similar fashion, in which it is again useful to transform to center of mass and relative coordinates and assume the conserved center of mass momentum $\vec{K}$ to be equal to zero. When the extra electron has $s^{e_2}\tau^{e_2}=s^{e_1}\tau^{e_1}=s^h\tau^h$, meaning that it can be excited simultaneously with the other electron and the hole, we find, when going to position representation, the decoupled differential equation
\begin{equation}
\label{tridif}
\begin{split}
\bigg(&-\mathcal{V}\left(E^{tri}_{\beta},r_{e_1h},r_{he_2}\right)\left(\nabla^2_{\vec{r}_{e_1h}}+\nabla^2_{\vec{r}_{he_2}}-\vec{\nabla}_{\vec{r}_{e_1h}}.\vec{\nabla}_{\vec{r}_{he_2}}\right)-\left(\frac{\partial}{\partial r_{e_1h}}\mathcal{V}\left(E^{tri}_{\beta},r_{e_1h},r_{he_2}\right)\right)\frac{\partial}{\partial r_{e_1h}} \\
&-\left(\frac{\partial}{\partial r_{he_2}}\mathcal{V}\left(E^{tri}_{\beta},r_{e_1h},r_{he_2}\right)\right)\frac{\partial}{\partial r_{he_2}}+\frac{1}{2}\left(\frac{\partial}{\partial r_{e_1h}}\mathcal{V}\left(E^{tri}_{\beta},r_{e_1h},r_{he_2}\right)\right)\frac{\partial}{\partial r_{he_2}}+\frac{1}{2}\left(\frac{\partial}{\partial r_{he_2}}\mathcal{V}\left(E^{tri}_{\beta},r_{e_1h},r_{he_2}\right)\right)\frac{\partial}{\partial r_{e_1h}} \\
&-V(r_{e_1h})-V(r_{he_2})+V(r_{e_1h}+r_{he_2})+\Delta_{s^h,\tau^h}+\frac{\Delta}{2}\bigg)\phi_{c,v,c}^{e_1,h,e_2}(r_{e_1h},r_{he_2}) = E^{tri}_{\beta}\phi_{c,v,c}^{e_1,h,e_2}(r_{e_1h},r_{he_2}),
\end{split}
\end{equation}
with
\begin{equation}
\label{Vdef}
\mathcal{V}\left(E^{tri}_{\beta},r_{e_1h},r_{he_2}\right) = \frac{2a^2t^2}{E^{tri}_{\beta}+V(r_{e_1h})+V(r_{he_2})-V(r_{e_1h}+r_{he_2})-\frac{\Delta}{2}}
\end{equation}
and where $\beta$ is a shorthand notation for $\alpha,s^{e_2},\tau^{e_2}$.
\end{widetext}

\end{document}